\documentclass[useAMS,usenatbib,onecolumn]{mn2e}

\def\spose#1{\hbox to 0pt{#1\hss}}
\def\simlt{\mathrel{\spose{\lower 3pt\hbox{$\mathchar"218$}}
     \raise 2.0pt\hbox{$\mathchar"13C$}}}
\def\simgt{\mathrel{\spose{\lower 3pt\hbox{$\mathchar"218$}}
     \raise 2.0pt\hbox{$\mathchar"13E$}}}
\def\simpropto{\mathrel{\spose{\lower 3pt\hbox{$\mathchar"218$}}
     \raise 2.0pt\hbox{$\propto$}}}

\usepackage{amsmath,graphicx,bm}
\title{Precision Calibration of Radio Interferometers Using Redundant Baselines}
\author[Adrian Liu, Max Tegmark, Scott Morrison, Andrew Lutomirski, Matias Zaldarriaga]{Adrian Liu$^{1}$\thanks{E-mail:
acliu@mit.edu}, Max Tegmark$^{1}$, Scott Morrison$^{1}$,Andrew Lutomirski$^{1}$, Matias Zaldarriaga$^{2}$\\
$^{1}$Dept. of Physics and MIT Kavli Institute, Massachusetts Institute of Technology, 77 Massachusetts Ave., Cambridge, MA 02139, USA\\
$^{2}$Institute for Advanced Study, School of Natural Sciences, Einstein Drive, Princeton, NJ 08540, USA}
\date{\today}
\begin{document}

\pagerange{\pageref{firstpage}--\pageref{lastpage}} \pubyear{2010}

\maketitle

\begin{abstract}
Growing interest in $21\,\textrm{cm}$ tomography has led to the design and construction of broadband radio interferometers with low noise, moderate angular resolution, high spectral resolution, and wide fields of view.  With characteristics somewhat different from traditional radio instruments, these interferometers may require new calibration techniques in order to reach their design sensitivities.  Self-calibration or redundant calibration techniques that allow an instrument to be calibrated off complicated sky emission structures are ideal.  In particular, the large number of redundant baselines possessed by these new instruments makes redundant calibration an especially attractive option.  In this paper, we explore the errors and biases in existing redundant calibration schemes through simulations, and show how statistical biases can be eliminated.  We also develop a general calibration formalism that includes both redundant baseline methods and basic point source calibration methods as special cases, and show how slight deviations from perfect redundancy and coplanarity can be taken into account.
\end{abstract}

\section{Introduction}

Technological advances of recent years have enabled the design and construction of radio telescopes with broadband frequency coverage, low instrumental noise, moderate angular resolution, high spectral resolution, and wide fields of view.  These instruments have generated considerable scientific interest due to their unique potential 
to do 21~cm cosmology, mapping the high-redshift universe using redshifted radio emission from neutral hydrogen.
Not only would this provide the only direct measurements of the cosmological dark ages and the Epoch of Reionization 
\citep{Rees, Tozzi2, Tozzi, Iliev,furlanetto1,Loeb1,furlanetto2, Barkana1,Whitepaper1}, but it also has the potential to overtake the cosmic microwave background as the most powerful cosmological probe, measuring cosmological parameters related to inflation, dark matter and neutrinos to unprecedented accuracy \citep{Matt3,Santos2, juddjackiemiguel1,Yi,Whitepaper2}
and improving provide dark energy constraints after reionization through the detection of baryon acoustic oscillations \citep{wyithe2008,ChangDE,miguelreview}.
 
From a practical standpoint, however, none of the aforementioned applications will be feasible without reliable algorithms for instrumental calibration.  Compared to traditional radio astronomy experiments, instruments designed for $21\,\textrm{cm}$ tomography will require particularly precise calibration because of the immense foreground subtraction challenges involved.  Epoch of Reionization experiments, for example, will be attempting to extract cosmological signals from beneath foregrounds that are of order $10^4$ times brighter \citep{angelicaGSM}.  Such procedures, where one must subtract strong foregrounds from a bright sky to uncover weak cosmological signals, are particularly susceptible to calibration errors.  A prime example of this is the effect that a miscalibration has on the subtraction of bright, resolved point sources.  \citet{Datta}, for instance, argued that bright point sources must be localized to extremely high positional accuracy in order for their subtraction to be successful, and \citet{us} showed that failure to do so could comprise one's subsequent ability to subtract any \emph{unresolved} point sources, making the detection of the $21\,\textrm{cm}$ cosmological signal difficult.  This places stringent requirements on calibration accuracy.

While standard algorithms are available for calibrating radio interferometers, the unprecedented challenges faced by $21\,\textrm{cm}$ tomography and the advent of a new class of large radio arrays (such as PAPER \citep{PAPER}, LOFAR \citep{LOFAR}, MWA \citep{MWAdesign}, CRT \citep{CRT}, the Omniscope \citep{FFTT2}, and the SKA \citep{SKA}) mean that it is timely to consider the relative merits of various algorithms.  For example, several authors \citep{direcdep,softwareholography} have pointed out that these new high precision radio arrays require close attention to be paid to heterogeneities in the primary beams of different antenna elements.  In this paper, we focus on the technique of \emph{redundant calibration}, which takes advantage of the fact that these new telescopes generically possess a large number of redundant or near-redundant baselines.  This property of exact or near-redundancy is not an accident, and arises from the science requirements of $21\,\textrm{cm}$ tomography\footnote{Note that we are \emph{not} advocating the redesign of arrays to achieve redundancy for the sake of redundant baseline calibration.  The design of an array should be driven by its science requirements, and the ability to use redundant calibration should simply be viewed as an option to be explored if there happens to be redundancy.}, where the demanding need for high sensitivity on large spatial scales calls for arrays with a large number of short baselines \citep{anothermiguel,adam}.  Since these arrays typically also have very wide fields-of-view and in many cases operate at frequencies where all-sky survey data is scarce, the fewer assumptions one needs to make about the structure of the sky signal, the better.  This is an advantage that redundant calibration has over other calibration methods --- no \emph{a priori} assumptions need to be made about calibrator sources.

The use of redundant baselines for calibration goes far back. For example, \citet{oldredund1}
used this approach to produce high dynamic range images of $3C84$, and similar techniques have been used in Solar imaging \citep{oldredund2,oldredund3}.  The methods used are described in detail in \citet{oldredund4,oldredund5,oldredund6}, and closely resemble the logarithmic implementation we review in Section \ref{log}.  In this paper, we build on this previous work and extend it in several ways:
\begin{enumerate}
\item We provide a detailed examination of the errors and biases involved in existing redundant baseline methods.
\item We introduce methods to mitigate (or, in the case of biases, eliminate) these problems.
\item We show how slight deviations from prefect redundancy and coplanarity can be incorporated into the analysis,
\end{enumerate}
with the view that it may be necessary to correct for all these effects in order to achieve the precision calibration needed for precision cosmology.

The rest of the paper is organized as follows.  After defining our notation, we describe the simplest possible calibration algorithm (\emph{i.e.} point source calibration) in Section \ref{traditional}.  In Section \ref{log} we review the redundant baseline calibration proposed by \citet{oldredund4}.  Readers familiar with radio interferometer calibration techniques are advised to skip to Section \ref{noisecovarsubsection}, where we examine Wieringa's scheme more closely by deriving the optimal weightings for its fitting procedure.  In Section \ref{simsubsection} we focus on simulations of redundant calibration for monolithic square arrays, which show that existing redundant calibration algorithms have certain undesirable error properties.  These properties suggest a linearized variation that we propose in Section \ref{linear}.  In Section \ref{generalized} we consider redundant calibration algorithms and point source calibration schemes as extreme cases of a generalized calibration formalism, which allows us to treat the case of near-redundant baseline calibration.  We summarize our conclusions in Section \ref{conc}. 

\section{Calibration Using Redundant Baselines}
\label{basic}

We begin by defining some basic notation.  Suppose an antenna $i$ measures a signal $s_i$ at a  given instant\footnote{More precisely, $s_i$ refers to frequency domain data at a given instant.  In other words, it is the result of dividing the time-ordered data into several blocks, each of which is then Fourier transformed in the time direction to give a signal $s_i$ that is both a function of frequency and time (albeit at a coarser time resolution than the original time-ordered data).}.  This signal can be written in terms of a complex gain factor $g_i$, the antenna's instrumental noise contribution $n_i$, and the true sky signal $x_i$ that would be measured in the limit of perfect gain and no noise:
\begin{equation}
s_i = g_i x_i + n_i.
\end{equation}
Each baseline measures the correlation between the two signals from the two participating antennas:
\begin{subequations}
\begin{eqnarray}
c_{ij} &\equiv & \langle s_i^{*} s_j \rangle \\
\label{zeromean} &=&  g_i^* g_j \langle x_i^* x_j \rangle +\langle n_i^* n_j \rangle+ g_i^* \langle x_i^* n_j \rangle + g_j \langle n_i^* x_j \rangle \quad \\
&=& g_i^* g_j \langle x_i^* x_j \rangle + n_{ij}^{res} \equiv g_i^* g_j y_{i-j} + n_{ij}^{res}, \label{measurement}
\end{eqnarray}
\end{subequations}
where we have denoted the true correlation $\langle x_i^* x_j \rangle$ by $y_{i-j}$, and the angled brackets $\langle \dots \rangle$ denote time averages.  The index $i-j$ is not to be taken literally but is intended to signify the fact that any given correlation should depend only on the \emph{relative} positions of the two antennas.  As is usual in radio interferometry, we have assumed that noise contributions from different antennas are uncorrelated, and that instrumental noise is uncorrelated with the sky signal.  This means that $\langle n_i^* x_j \rangle$ reduces to $\langle n_i^* \rangle \langle x_j \rangle$, and since $\langle n_i \rangle \rightarrow 0$ in the limit of long integration times, we can assume that the last three terms in Equation \ref{zeromean} average to some residual noise level $n^{res}$.  The size of $n^{res}$ will of course depend on the details of the instrument, but in general will be on the order of $T_{sys} / \sqrt{\tau \Delta \nu}$, where $T_{sys}$ is some fiducial system temperature, $\tau$ is the total integration time, and $\Delta \nu$ is the bandwidth.  For the purposes of this paper, the details of the noise term do not matter, and for notational convenience we will drop the superscript ``res" from now on.  The key qualitative results of this paper (such as the fact that many existing redundant baseline calibration schemes are biased) do not depend on the noise level.

It should be noted that the equation for $c_{ij}$ presented above does not represent the most general form possible for the measured correlations.  So-called non-closing errors, for example, can result in gain factors that do not factor neatly on an antenna-by-antenna basis.  In this paper we will forgo a discussion of these errors, with the view that a good number of these contributions can be mitigated (although not eliminated completely) by good hardware design.  Alternatively, our results may be interpreted as best-case scenario \emph{upper limits} on calibration quality.  

Since the gain factors in Equation \ref{measurement} are in general complex, we can parameterize them by an amplitude gain $\eta$ and a phase $\varphi$ by letting $g_i \equiv e^{\eta_i + i \varphi_i}$.  Our equation then becomes
\begin{equation}
\label{toomanyvars}
c_{ij} = \exp \left[ (\eta_i + \eta_j) + i (\varphi_j-\varphi_i) \right] y_{i-j}+n_{ij}.
\end{equation}
The goal of any radio interferometer is to extract the true correlations $y_{i-j}$ (from which one can construct a sky map using Fourier transforms) from the measured correlations $c_{ij}$.  Formally, however, this is an unsolvable problem for the generic case, as one can see from Equation \ref{toomanyvars} --- with $N$ antennas, one has $N(N-1)/2$ measured correlations ($c_{ij}$'s), which are insufficient to solve for the $N(N-1)/2$ true correlations ($y_{i-j}$'s) in addition to the $2N$ unknown $\eta$'s and $\varphi$'s.  (The $n_{ij}$ noise terms are also unknown quantities, but are generally not treated as ones that need to be solved for, but as noise to be minimized in a least-squares fit).  In abstract terms, all calibration schemes can be thought of as methods for reducing the number of unknowns on the right hand side of Equation \ref{toomanyvars} so that the system of equations becomes overdetermined and the $\eta$'s and $\varphi$'s can be solved (or fit) for.
\subsection{Basic Point Source Calibration}
\label{traditional}
The most straightforward way to calibrate a radio telescope is to image a single bright point source.  A point source has constant amplitude in $uv$-space, and thus $y_{i-j}$ is a complex number whose complex amplitude is independent of baseline.  Moreover, the phase of $y_{i-j}$ must also be zero, since by definition we are only ``looking" at the bright point source if our radio array is phased so that the phase center lies on the source.  Thus, $y_{i-j} = S$, and Equation \ref{toomanyvars} reduces to
\begin{equation}
\label{trad}
c_{ij} = \exp \left[ (\eta_i + \eta_j) + i (\varphi_j-\varphi_i) \right] S+n_{ij},
\end{equation}
which is an overdetermined set of equations for all but the smallest arrays, since we have $N(N-1)/2$ complex inputs on the left hand side but only $2N$ real numbers ($\eta$'s and $\varphi$'s) to solve for on the right hand side.  The system is therefore solvable, up to two intrinsic degeneracies: the overall gain $\sum_i \eta_i$ is indeterminate, as is the overall phase $\sum_i \varphi_i$.  These degeneracies, however, can be easily dealt with by having some knowledge of our instrument and our calibration source.  For instance, the overall gain can be computed if the calibrator source is a catalogued source with known flux density.

It is important to note that the point source calibration method we have presented in this section represents only the simplest, most straightforward way in which one could calibrate a radio telescope.  
There exist far more sophisticated methods\footnote{See \citet{review1} or \citet{review2}, for example, for nice reviews.}, such as the many self-calibration schemes that are capable of calibrating a radio telescope off most reasonable sky signals.  In practice, one almost never needs to use basic point source calibration, for it is just as easy to use a self-calibration algorithm.  The scheme described here should therefore be considered as no more than a ``toy model" for calibration.

\subsection{Redundant Baseline Calibration}
\label{our}
The main drawback of the approach described in the previous section is that it requires the existence of an isolated bright point source.  This requirement becomes increasingly difficult to fulfill as cosmological science drivers push radio telescopes towards regimes of higher sensitivity and wider fields of view.  While self calibration methods have been shown to produce high dynamic range maps over wide fields of view and do \emph{not} require the existence of isolated bright point sources, they are reliant on having a reasonable \emph{a priori} model of the sky.  Thus, if possible it may be preferable to opt for redundant calibration, which is completely model independent.  With the emergence of telescopes with high levels of redundancy (such as the Cylinder Radio Telescope --- see \citet{CRT} --- or any omniscopes\footnote{We define an omniscope as any instrument where antenna elements are arranged on a regular grid, and full digitized data is collected on a per element basis without any beamforming.} like those described in \citet{FFTT1,FFTT2}) redundant baseline calibration becomes a competitive possibility.

Mathematically, if an array has a large number of redundant baselines, then one can reduce the number of unknowns on the right hand side of Equation \ref{toomanyvars} by demanding that the true visibilities $y_{i-j}$ for a set of identical baselines be the same.  Since the number of \emph{measured} visibilities $c_{ij}$'s stays the same, our system of equations becomes overdetermined provided there are a sufficient number of redundant baselines, and it becomes possible to fit for the $\eta$ and $\varphi$ parameters.

As an example, consider a one-dimensional array of five radio antennas where the antennas are equally spaced.  In this case, one has four unique baselines (of lengths 1, 2, 3, and 4), and 10 measured correlations.  Thus, one must fit for 18 real numbers (four complex numbers from the true visibilities of the four unique baselines, plus five $\eta$'s and five $\varphi$'s) from 20 numbers (10 complex measured correlations):
\begin{eqnarray}
\nonumber
c_{1,2} &=& \exp \left[ (\eta_1 + \eta_2) + i (\varphi_2-\varphi_1) \right] y_{1} +n_{1,2}\\ 
\nonumber
c_{2,3} &=& \exp \left[ (\eta_2 + \eta_3) + i (\varphi_3-\varphi_2) \right] y_{1} +n_{2,3}\\
\nonumber
& \vdots & \textrm{(all four baselines of length 1)}\\
\nonumber
c_{1,3} &=& \exp \left[ (\eta_1 + \eta_3) + i (\varphi_3-\varphi_1) \right] y_{2} +n_{1,3}\\
\label{setofeqns}
c_{2,4} &=& \exp \left[ (\eta_2 + \eta_4) + i (\varphi_4-\varphi_2) \right] y_{2} +n_{2,4}\\
\nonumber
& \vdots & \textrm{(all three baselines of length 2)}\\
\nonumber
c_{1,4} &=& \exp \left[ (\eta_1 + \eta_4) + i (\varphi_4-\varphi_1) \right] y_{3} +n_{1,4}\\
\nonumber
& \vdots & \textrm{(both baselines of length 3)} \\
\nonumber
c_{1,5} &=& \exp \left[ (\eta_1 + \eta_5) + i (\varphi_5-\varphi_1) \right] y_{4}+n_{1,5}.
\end{eqnarray}
As is usual in radio interferometry, we have omitted equations corresponding to baselines of length zero, \emph{i.e.} autocorrelations of the data from a single antenna.  This is because autocorrelation measurements carry with them correlated noise terms, and thus are likely to be much noiser than other correlated measurements.

In the following sections we will present two methods for explicitly solving Equation \ref{setofeqns}.  We begin in Section \ref{log} with a logarithmic method, which is very similar to one described by  \citet{oldredund4}.  This method is the simplest way to implement a redundant baseline calibration scheme, but suffers from several problems which we detail in Section \ref{simsubsection} and \ref{logprobs}. In Section \ref{linear}, we show how these problems can be
solved with an alternative method based on linearization.

\subsection{The Logarithmic Implementation}
\label{log}
We proceed by rewriting our equations in the following way:
\begin{equation}
\label{prelog}
c_{ij}=g_i^{*}g_j y_{i-j} \left( 1+ \frac{n_{ij}}{g_i^* g_j y_{i-j}}\right).
\end{equation}
Taking the logarithm gives
\begin{equation}
\ln c_{ij} = \eta_i + \eta_j + i ( \varphi_j - \varphi_i ) + \ln y_{i-j} + \underbrace{\ln \left( 1+ \frac{n_{ij}}{g_i^* g_j y_{i-j}}\right)}_{\equiv w_{ij}},
\end{equation}
and requiring that the real and imaginary parts be separately equal gives linear equations of the form
\begin{subequations}
\begin{eqnarray}
\label{logreal}
\ln \left| c_{i,j} \right| &=& \eta_i + \eta_j + \ln \left| y_{i-j} \right| + \textrm{Re} \, w_{ij}\\
\label{logim}
\arg \left| c_{i,j} \right| &=& \varphi_j - \varphi_i + \arg \left| y_{i-j} \right| + \textrm{Im} \,w_{ij}.
\end{eqnarray}
\end{subequations}
Note that the amplitude and phase calibrations have now decoupled\footnote{From a rigorous standpoint, it is not strictly true that the amplitude and phase calibrations of an interferometer decouple.  Indeed, this is evident even in our simple model, where $\textrm{Re} \, w_{ij}$ and $\textrm{Im} \, w_{ij}$ both contain factors of the complex gain $g \equiv e^{\eta + i \varphi}$, which means that Equations \ref{logreal} and \ref{logim} are in principle coupled, even if only weakly so.  However, one must remember that the noise parameters are treated simply as given numbers which are \emph{not} solved for when solving these calibration equations.  This means that in finding the best fit solution to Equations \ref{logreal} and \ref{logim}, the $\eta$ and $\varphi$ calibrations \emph{can be performed separately}.  It is only in this sense that the systems are ``decoupled".}, and can thus be written as two separate matrix systems.  The amplitude calibration, for example, takes the form
\begin{equation}
\label{matrix1}
\underbrace{\left(\begin{array}{c}
\ln \left| c_{1,2} \right| \\
\ln \left| c_{2,3} \right| \\
\ln \left| c_{3,4} \right| \\
\ln \left| c_{4,5} \right| \\
\ln \left| c_{1,3} \right| \\
\ln \left| c_{2,4} \right| \\
\ln \left| c_{3,5} \right| \\
\vdots \\
\ln \left| c_{2,5} \right| \\
\ln \left| c_{1,5} \right|
\end{array} \right)}_{\equiv \mathbf{d}}=
\underbrace{\left( \begin{array}{ccccccccc}
1 & 1 & 0 & 0 & 0 & 1 & 0 & 0 & 0\\
0 & 1 & 1 & 0 & 0 & 1 & 0 & 0 & 0 \\
0 & 0 & 1 & 1 & 0 & 1 & 0 & 0 & 0 \\
0 & 0 & 0 & 1 & 1 & 1 & 0 & 0 & 0 \\
1 & 0 & 1 & 0 & 0 & 0 & 1 & 0 & 0 \\
0 & 1 & 0 & 1 & 0 & 1 & 1 & 0 & 0 \\
0 & 0 & 1 & 0 & 1 & 1 & 1 & 0 & 0 \\
& \vdots & &  & \vdots & & & \vdots & \\
0 & 1 & 0 & 0 & 1 & 0 & 0 & 1 & 0 \\
1 & 0 & 0 & 0 & 1 & 0 & 0 & 0 & 1
\end{array} \right)}_{\equiv \mathbf{A}}
\underbrace{\left(\begin{array}{c}
\eta_1 \\
\eta_2 \\
\eta_3 \\
\eta_4 \\
\eta_5 \\
\ln \left| y_1 \right| \\
\ln \left| y_2 \right| \\
\ln \left| y_3 \right| \\
\ln \left| y_4 \right|
\end{array} \right)}_{\equiv \mathbf{x}}+
\underbrace{\left(\begin{array}{c}
s_{1,2} \\
 s_{2,3} \\
 s_{3,4} \\
s_{4,5} \\
 s_{1,3} \\
 s_{2,4} \\
 s_{3,5} \\
\vdots \\
 s_{2,5} \\
 s_{1,5}
\end{array} \right)}_{\equiv \textrm{Re} \,\mathbf{w}},
\end{equation}
for the one-dimensional, five-element array considered in Section \ref{our}.  The vector $\mathbf{d}$ stores the measured correlations, the matrix $\mathbf{A}$ is completely determined by the array configuration, and the $\mathbf{x}$ is what we hope to solve for.

As it currently stands, however, this set of linear equations has no unique solution even in the absence of noise (\emph{i.e.} even when $\mathbf{w}=0$).  This can be seen from the fact that the vector $\mathbf{x}_{null} = \left( 1, 1,1,1,1,-2,-2,-2,-2 \right)$ (written here as a row vector for convenience) lies in the null space of $\mathbf{A}$ (\emph{i.e.} $\mathbf{A}\mathbf{x}_{null}=0$), so for any solution $\mathbf{x}_0$, one can form a new solution $\mathbf{x}_0 + \mathbf{x}_{null}$.  The new solution corresponds to adding a constant to all $\eta$'s, which --- since $g \equiv \exp(\eta + i \varphi)$ --- is equivalent to multiplying all gains by a constant factor and simultaneously dividing all the sky signals by the same factor.  Physically, this is an expression of the fact that internal calibration schemes (\emph{i.e.} schemes like this one where the calibration is done by exploiting internal mathematical consistencies) are insensitive to the absolute gain of the radio interferometer as a whole, a problem which is present even performing a simple point source calibration, as we noted in Section \ref{traditional}.

This mathematical degeneracy in absolute amplitude calibration can be broken by arbitrarily specifying an absolute gain level.  For example, one can add the equation
\begin{equation}
\label{degen1}
0 = \sum_i \eta_i
\end{equation}
to the system of equations, thus ensuring that one cannot uniformly elevate the gain levels and still satisfy all constraints.  The set of equations governing the phase calibration require the addition of a similar equation, since the absolute phase of a system is not a physical meaningful quantity and therefore must be arbitrarily specified:
\begin{equation}
0 = \sum_i \varphi_i.
\end{equation}
With the phase calibration, however, there exist two additional degeneracies: the calibration is insensitive to tilts of the entire telescope in either the $x$ or the $y$ direction, since such tilts are equivalent to rotations of the sky and redundant algorithms are (by design) independent of the nature of the sky signal.  To break these degeneracies, one adds the following equations:
\begin{subequations}
\begin{eqnarray}
0 &=& \sum_i r_{x,i} \varphi_i \\
0 &=& \sum_i r_{y,i} \varphi_j,
\end{eqnarray}
\end{subequations}
where $\mathbf{r}_i \equiv (r_{x,i}, r_{y,i} )$ denotes the physical location of the $i$th antenna in the array.  While these extra equations are somewhat arbitrary (the L.H.S. of Equation \ref{degen1}, for instance, can be any real number), the true gain, phase, and tilt parameters can always be fixed after-the-fact by referring to published flux densities and locations of known, catalogued bright point sources in the final sky maps.

With these four degeneracies broken, our equations can be solved using the familiar least-squares estimator $\mathbf{\hat{x}}$ for the parameter vector:
\begin{equation}
\label{least^2}
\mathbf{\hat{x}} =\mathbf{ \left[ A^t N^{-1} A \right]^{-1} A^t N^{-1}} \mathbf{d}.
\end{equation}
Here, $\mathbf{N}$ is the noise covariance matrix, which takes the form $\langle \textrm{Re}\, \mathbf{w} \,\textrm{Re} \,\mathbf{w^t} \rangle$ for the amplitude calibration and $\langle \textrm{Im}\, \mathbf{w} \,\textrm{Im} \,\mathbf{w^t} \rangle$ for the phase calibration.  In the following section we examine the form of this matrix in detail.  Note that since the parameter vector $\mathbf{x}$ contains not only the calibration factors $\eta$ and $\varphi$ but also the true sky correlations $y_{i-j}$, one can solve for the true sky signal directly from the uncalibrated correlations $\mathbf{d}$, as one expects from any self or redundant calibration scheme.

This concludes our review of the \citet{oldredund4} redundant calibration method, cast in notation conducive to the development of the new material that follows.  It is worth noting that in \citet{oldredund4}, the noise covariance matrix $\mathbf{N}$ is set to the identity.  In the next section, we show that this is \emph{not} the optimal weighting to use for the calibration fit, and derive the noise covariance matrix that corresponds to inverse variance weighting.

\subsection{The Noise Covariance Matrix}
\label{noisecovarsubsection}
The form of $\mathbf{N}$ will depend on one's model for the instrumental noise.  In what follows, we will for convenience assume that the noise is Gaussian, small compared to the signal, and uncorrelated between baselines; generalization to other noise models is straightforward\footnote{Typically, the electronic noise will be uncorrelated between different antennas/amplifier chains. This means that the noise correlation matrix $\mathbf{N}$ for the baseline will be highly sparse but not diagonal: away from its diagonal, it will have nonzero entries for precisely those pairs of baselines which have an antenna in common. This sparseness is very helpful, making the computations numerically feasible even for massive arrays as described below.}.  
Although these assumptions are not necessary for calibration (since Equation \ref{least^2} can be implemented for any invertible choice of $\mathbf{N}$), making them will allow us to gain a more intuitive understanding of the errors involved.  Recalling the form of the noise term after taking the logarithm of our equations, we can say
\begin{equation}
\label{thing1}
w_{ij} = \ln \left( 1+ \frac{n_{ij}}{g_i^* g_j y_{i-j}}\right) \approx  \frac{n_{ij}}{g_i^* g_j y_{i-j}},
\end{equation}
where we have invoked the assumption of high signal ($g_i^* g_j y_{i-j}$) to noise ($n_{ij}$) to expand the logarithm.  Since $c_{ij} = g_i^* g_j y_{i-j} +n_{ij}$, we can rewrite this as
\begin{equation}
\label{thing2}
w_{ij} \approx \frac{n_{ij}}{c_{ij} - n_{ij}} \approx \frac{n_{ij}}{c_{ij} },
\end{equation}
where we have neglected higher order terms.  Equation \ref{thing2} is more convenient than Equation \ref{thing1} because it is written in terms of the measured correlation $c_{ij}$ instead of the noiseless $g_i^* g_j y_{i-j}$.  Since the logarithmic scheme separates the real and imaginary parts of the calibration, the crucial quantity is not $w_{ij}$ but rather its real (or imaginary) part:
\begin{subequations}
\begin{eqnarray}
w_{ij}  &\approx & \frac{n_{ij}}{c_{ij} } = \frac{e^{- i \phi}}{|c_{ij}|}(n_x + i n_y) \\
&=&  \frac{1}{|c_{ij}|}(\cos \phi - i \sin \phi)(n_x + i n_y) \\
\Rightarrow \textrm{Re} \, w_{ij} &=& \frac{1}{|c_{ij}|} ( n_x \cos \phi + n_y \sin \phi),
\end{eqnarray}
\end{subequations}
where $\phi \equiv \textrm{arg\,}c_{ij}$.  With this result, we can form the noise covariance matrix $\mathbf{N} \equiv \langle \textrm{Re}\, \mathbf{w} \,\textrm{Re} \,\mathbf{w}^t \rangle$.  Since we are assuming that the noise is uncorrelated between different baselines, our matrix must be diagonal, and its components can be written in the form $\mathbf{N}_{\alpha \beta} = \langle (\textrm{Re} \, \mathbf{w}_{\alpha} )^2 \rangle \delta_{\alpha \beta}$ where $\delta_{\alpha \beta}$ is the Kronecker delta, and $\alpha$ and $\beta$ are Greek \emph{baseline} indices formed from \emph{pairs} of Latin \emph{antenna} indices.  For example, the baseline formed by antennas $i=1$ and $j=2$ might be labeled baseline $\alpha=1$, whereas that formed by antennas $i=1$ and $j=3$ might be given baseline index $\alpha=2$.  The diagonal components are given by
\begin{subequations}
\begin{eqnarray}
\mathbf{N}_{\alpha \alpha}&=& \langle n_{\alpha} n_{\alpha} \rangle \\
&=& \frac{1}{|c_{\alpha}|^2} ( \underbrace{\langle n_x n_x \rangle}_{\equiv \sigma^2} \cos^2 \phi + 2 \underbrace{\langle n_x n_y \rangle}_{=0} \cos \phi \sin \phi +\underbrace{\langle n_y n_y \rangle}_{\equiv \sigma^2} \sin^2 \phi ) \label{nxny} \\
&=& \frac{\sigma^2}{|c_{\alpha}|^2},
\label{Ncovar}
\end{eqnarray}
\end{subequations}
where we have assumed that the real and imaginary parts of the residual noise $n$ are independently Gaussian distributed with spread $\sigma^2$.  This expression provides a simple prescription for the inverse variance weighting of Equation \ref{least^2}: one simply weights each measured correlation by the square modulus of itself (the $\sigma^2$ in the numerator is irrelevant in the computation of the estimator $\mathbf{\hat{x}}$ because the two factors of $\mathbf{N}$ in Equation \ref{least^2} ensure that any constants of proportionality cancel out).

\begin{figure}
\centering
\includegraphics[width=0.5\textwidth,trim=1.5cm 5.0cm 1.5cm 5.0cm, clip]{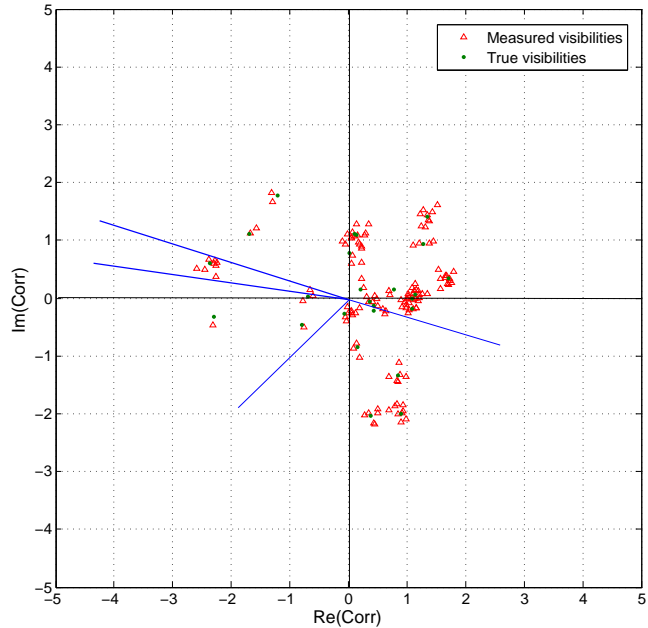}
\caption{A plot of true visibilities (green dots) and measured noisy visibilities (red triangles) on the complex plane.  The blue lines delineate the phase extent of the noisy visibilities from two sets of redundant baselines.  One sees that in the presence of noise, visibilities closer to the origin are more susceptible to loss of phase information.}
\label{phaseloss}
\end{figure}

For the phase calibration, the fact that inverse variance weighting corresponds to weighting by $1/|c|^2$ has a simple interpretation.  In Figure \ref{phaseloss}, we plot various correlations on the complex plane.  The green dots represent a set of simulated noiseless visibilities, while the red triangles show what the baselines of a noisy 4 by 4 regular square antenna array would measure.  From the plot, it is clear that for a given amount of noise, visibilities that are closer to the origin of the complex plane (\emph{i.e.} those with smaller $|c|$) are more susceptible to loss of phase ($\textrm{arg\,}c$) information from noise.  The gain ($\textrm{log\,}|c|$) errors are similarly large for such visibilities.  These visibilities should be given less weight in the overall fit, which is what the inverse variance weighting achieves.

\begin{figure}
\centering
\includegraphics[width=0.5\textwidth,trim=1.4cm 6.0cm 1.75cm 6.8cm, clip]{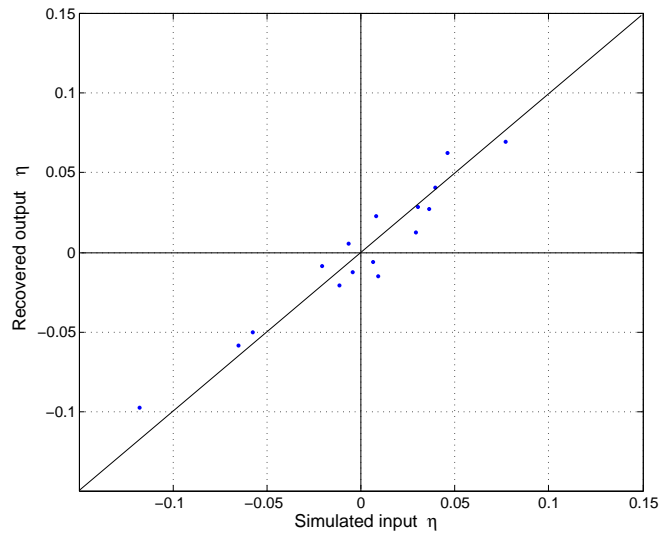}
\caption{Scatter plot of simulated vs. recovered antenna gain parameter $\eta$ for a 4 by 4 square array with a signal-to-noise ratio of 10.}
\label{logeta}
\end{figure}
\begin{figure}
\centering
\includegraphics[width=0.5\textwidth,trim=1.4cm 6.0cm 1.75cm 6.8cm, clip]{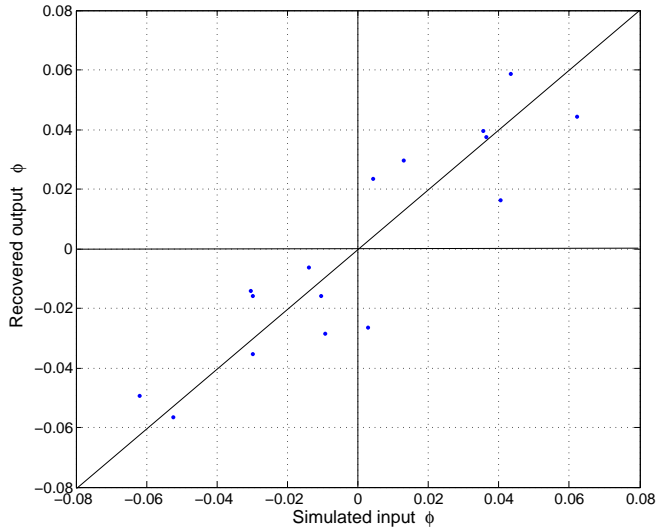}
\caption{Scatter plot of simulated vs. recovered antenna phase parameter $\varphi$ for a 4 by 4 square array with a signal-to-noise ratio of 10.}
\label{logphi}
\end{figure}
\subsection{Simulation Results and Error Properties}
\label{simsubsection}
In Figures \ref{logeta} and \ref{logphi}, we show simulated calibration results of a 16 antenna element interferometer.  The sky signal used in the simulations was that of a Gaussian random field.  That such a sky is somewhat unrealistic is irrelevant, since redundant calibration schemes are sky-independent\footnote{That is, provided we exclude unreasonable mathematical exceptions such as a sky devoid of any sources.} by construction.

Like in Figure \ref{phaseloss}, the antennas in these simulations were arranged in a 4 by 4 square grid, with the antenna spacings assumed to be completely regular.  The precise physical separation between adjacent antennas need not be specified, because a dilation of the entire array layout changes only the sky signal measured by the interferometer, and not the \emph{pattern} of baseline redundancies.  Repeating the simulations with rescaled versions of the same array is therefore unnecessary, again because of the fact that we can calibrate from any sky signal.  The same reasoning makes it unnecessary to specify an observing frequency for the simulations.

In the plots, it is clear that there is some scatter in the recovered antenna parameters.  This is due to non-zero instrumental noise, which was simulated to be Gaussian \emph{i.e.} $n_{ij}$ in Equation \ref{measurement} was taken to be a Gaussian random variable.  Rather than using a specific noise model to fix the amplitude of the noise, we simply parameterize the problem in terms of the signal-to-noise ratio (SNR).  Figures \ref{logeta} and \ref{logphi} show the results for an SNR of 10.  (Such an SNR would be typical for an interferometer with bandwidth $\Delta \nu \approx 10\,\textrm{kHz}$ and integration time $\tau \approx 10\,\textrm{s}$ observing Galactic synchrotron emission at $150\,\textrm{MHz}$).  The scatter seen in these figures can be viewed as error bars in the final calibration parameters outputted from the calibration.  To quantify these errors, we can compute the deviation $\mathbf{\varepsilon} \equiv \mathbf{x} - \mathbf{\hat{x}}$ and form the quantity
\begin{equation}
\label{errorbars}
\mathbf{\Sigma} \equiv \langle \mathbf{\varepsilon} \mathbf{\varepsilon}^t \rangle = \mathbf{\left[ A^t N^{-1} A \right]^{-1}},
\end{equation}
where the last equality can be proved with a little algebra and Equation \ref{least^2} \citep{MaxMaps}.  The square root of the diagonal elements of $\Sigma$ then give us the expected error bars on the corresponding elements in $\mathbf{\hat{x}}$.  For example, $\sigma_1$, the expected error in the calibration of the first antenna's $\eta$, is given by $(\mathbf{\Sigma}_{11})^{1/2}$ if one arranges the elements of the parameter vector $\mathbf{x}$ in the manner shown in Equation \ref{matrix1}.

The fact that our system is linear means that to understand the errors in our calibration, we need only compute error bars from simulations of systems with an SNR of 1; the errors from systems of arbitrary SNR can be scaled accordingly.  To see this, consider Equation \ref{Ncovar}, which we can rewrite by scaling out some fiducial mean value $|c_0|$ for the magnitude of the visibilities:
\begin{subequations}
\begin{eqnarray}
\mathbf{N}_{\alpha \beta} &=& \frac{\sigma^2}{|c_{\alpha}|^2} \delta_{\alpha \beta} = \frac{\sigma^2}{|c_0|^2 |c^{\prime}_{\alpha}|^2} \delta_{\alpha \beta} = \frac{1}{(\textrm{SNR})^2} \frac{\delta_{\alpha \beta}}{|c^{\prime}_{\alpha}|^2} \qquad \\
&\equiv & \frac{\mathbf{N}^{\prime}_{\alpha \beta}}{(\textrm{SNR})^2},
\end{eqnarray}
\end{subequations}
where we have defined normalized visibilities $c^{\prime}_\alpha$ and a normalized noise covariance matrix $\mathbf{N}^{\prime}\equiv  \frac{\delta_{\alpha \beta}}{|c^{\prime}_{\alpha}|^2}$.  Plugging this into Equation \ref{errorbars} gives
\begin{equation}
\label{rescaled error bars}
\mathbf{\Sigma} = \mathbf{\left[ A^t N^{-1} A \right]^{-1}}=\frac{1}{(\textrm{SNR})^2} \mathbf{\left[ A^t {N^{\prime}}^{-1} A \right]^{-1}}.
\end{equation}
We therefore see that in what follows, general conclusions for any SNR can be scaled from our simulations of a system with an SNR of 1.

Although redundant calibration schemes make it \emph{possible} to calibrate an interferometer using any reasonable sky signal, it is important to point out that the \emph{quality} of the calibration, or in other words, the error bars predicted by Equation \ref{rescaled error bars}, \emph{are} sky-dependent.  To see this, consider a situation where the sky signal is dominated by a small number of Fourier modes whose wavevectors are of precisely the right magnitude and orientation for its power to be picked up by only the longest baselines of an array.  Since our noise covariance matrix is proportional to $1/|c|$, using such a sky to calibrate our array will give small error bars for the antennas that are part of the longest baselines and relatively large error bars for all other antennas.  On the other hand, a sky that has comparable power across all scales is likely to yield calibration error bars that are more consistent across different antennas.  Given this dependence of calibration quality on sky signal, in what follows we compute ensemble-averaged error bars based on random realizations of our Gaussian skies to give a sense of the typically attainable accuracy.

\begin{figure}
\centering
\includegraphics[width=0.5\textwidth]{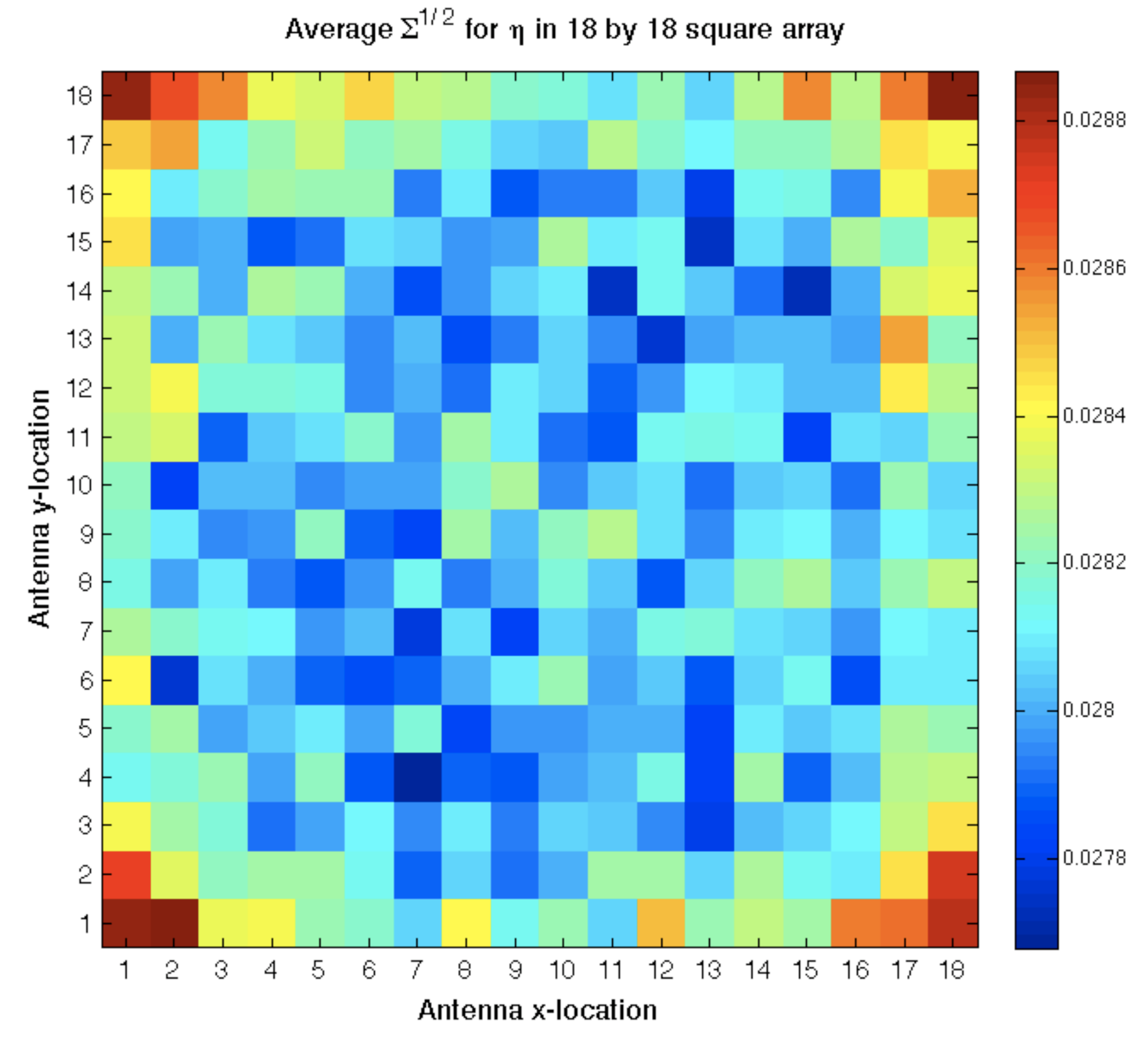}
\caption{Expected error bars in recovered $\eta$ for different antennas in an 18 by 18 square array with a signal-to-noise (SNR) ratio of 1, although in a realistic situation one would expect a more favorable SNR.}
\label{antennalayout}
\end{figure}

\begin{figure}
\centering
\includegraphics[width=0.5\textwidth,trim=1.0cm 5.0cm 1.0cm 4.0cm, clip]{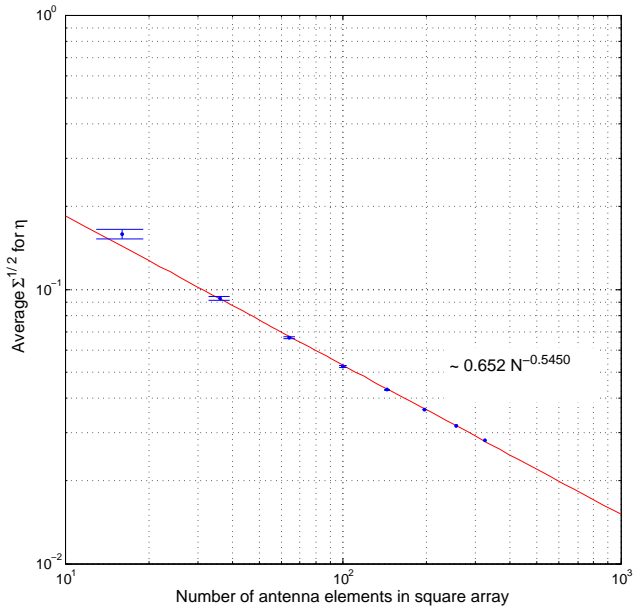}
\caption{Expected average error bars in recovered $\eta$ as a function of the size of the simulated square array.  All simulations have a signal-to-noise (SNR) ratio of 1, although in a realistic situation one would expect a more favorable SNR.}
\label{totalnumantennas}
\end{figure}

From our simulations, we find that the dependence of calibration quality on antenna location is quite weak.  This can be seen from the scale on Figure \ref{antennalayout}, where we show the expected error bars (estimated by taking an average of 30 random realizations) in the recovered $\eta$ for the antennas that comprise a regularly spaced 18 by 18 square interferometer.  Far more important than antenna location in determining the calibration errors is the \emph{total} number of antennas in an array.  In Figure \ref{totalnumantennas}, we show the average\footnote{Averaging both over different antennas in an array and over different simulations.}  error in $\eta$ as a function of the number of antennas in a square array.  The error $\Sigma^{1/2}$ roughly asymptotes to a $1/\sqrt{N}$ dependence, where $N$ is the total number of antennas in the array, as would be expected if the calibration algorithm is thought of as combining independent information from different redundant baselines.  Precisely the same trend is seen in the $\varphi$ errors.

In running the simulations, we find that minor variations in the baseline distribution have little effect on the final calibration errors\footnote{\emph{Major} variations, of course, can drastically affect the errors.  An extreme example of this would be a change in baseline distribution that completely destroys all redundancy, which would render the algorithm unusable and formally give rise to infinite error bars.}.  For instance, varying the aspect ratio of a 100 antenna system from a 10 by 10 array, to a 5 by 20, 4 by 25, and a 1 by 100 array has mostly minimal effects on the $\eta$ and $\varphi$ errors.  Thus, for an array with baseline redundancies that are at roughly the same level as for a square array, we can put everything together and arrive at the following rule of thumb for the calibration errors:
\begin{equation}
\label{neateqn}
\Delta \eta \equiv \Sigma_\eta^{1/2} \approx \Delta \varphi \equiv \Sigma_\varphi^{1/2} \approx\frac{0.5}{\textrm{SNR}} \frac{1}{\sqrt{N}},
\end{equation}
where $N$ is again the total number of antennas in the array.

Not captured by Equation \ref{neateqn} is the fact that antennas closer to the center of an array tend to have lower calibration errors than those that are peripherally located, as can be seen in Figure \ref{antennalayout}.  Intuitively, this is due to the fact that while every antenna forms the same number of baselines (because each antenna forms a baseline with all other antennas in the array), antennas that are located in different parts of the array form different \emph{sets} of baselines.  An antenna that is in the corner of a square array, for example, forms $N-1$ baselines that are all different from each other, whereas an antenna that is close to the center will form baselines in redundant pairs because of the symmetry of the square array.  In the system of equations, the centrally located antenna will therefore be more tightly constrained by the data, and the resulting error bars in its calibration parameters will be smaller.  These differences are, however, rather minimal and are unlikely to significantly affect the actual calibration of a real (non-simulated) array.

\begin{figure}
\centering
\includegraphics[width=0.5\textwidth]{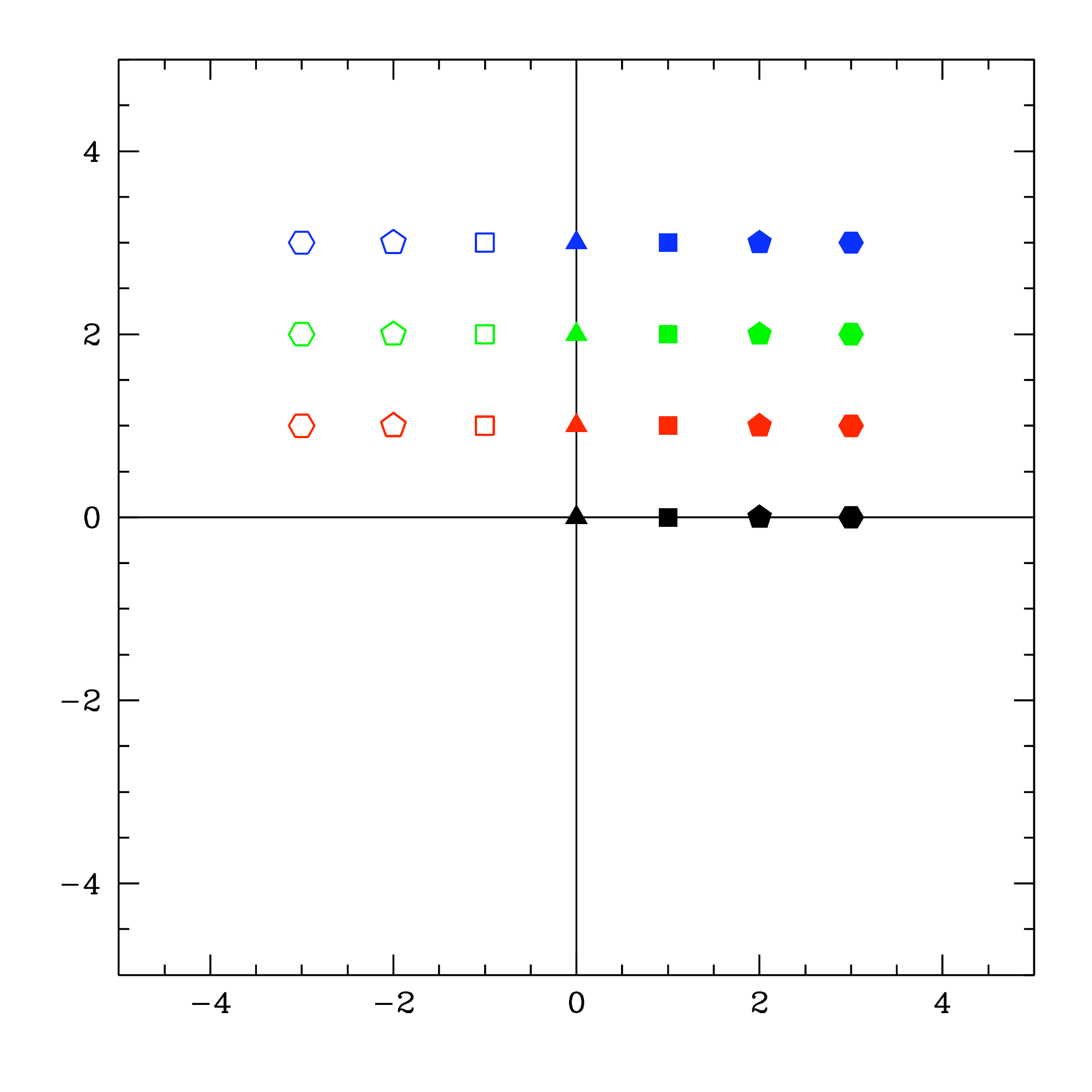}
\caption{Legend of baselines for Figures \ref{originalvisibilities} and \ref{adjustedvisibilities}.}
\label{legend}
\end{figure}

\begin{figure}
\centering
\includegraphics[width=0.5\textwidth]{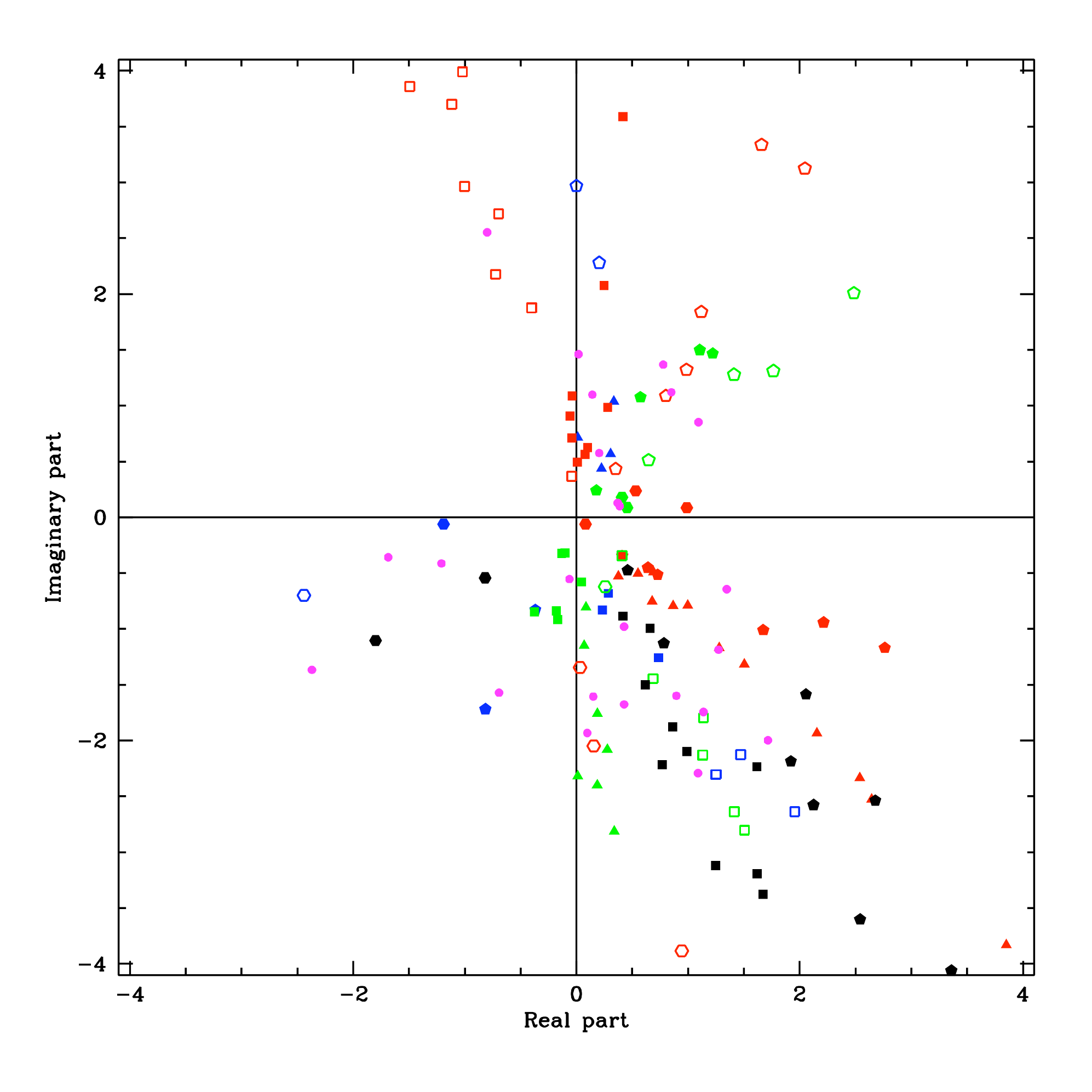}
\caption{Uncalibrated visibility measurements plotted on the complex plane according to the $uv$ plane baseline legend provided by Figure \ref{legend}.  The simulated (\emph{i.e.} true) sky visibilities are given by the magenta dots.}
\label{originalvisibilities}
\end{figure}

\begin{figure}
\centering
\includegraphics[width=0.5\textwidth]{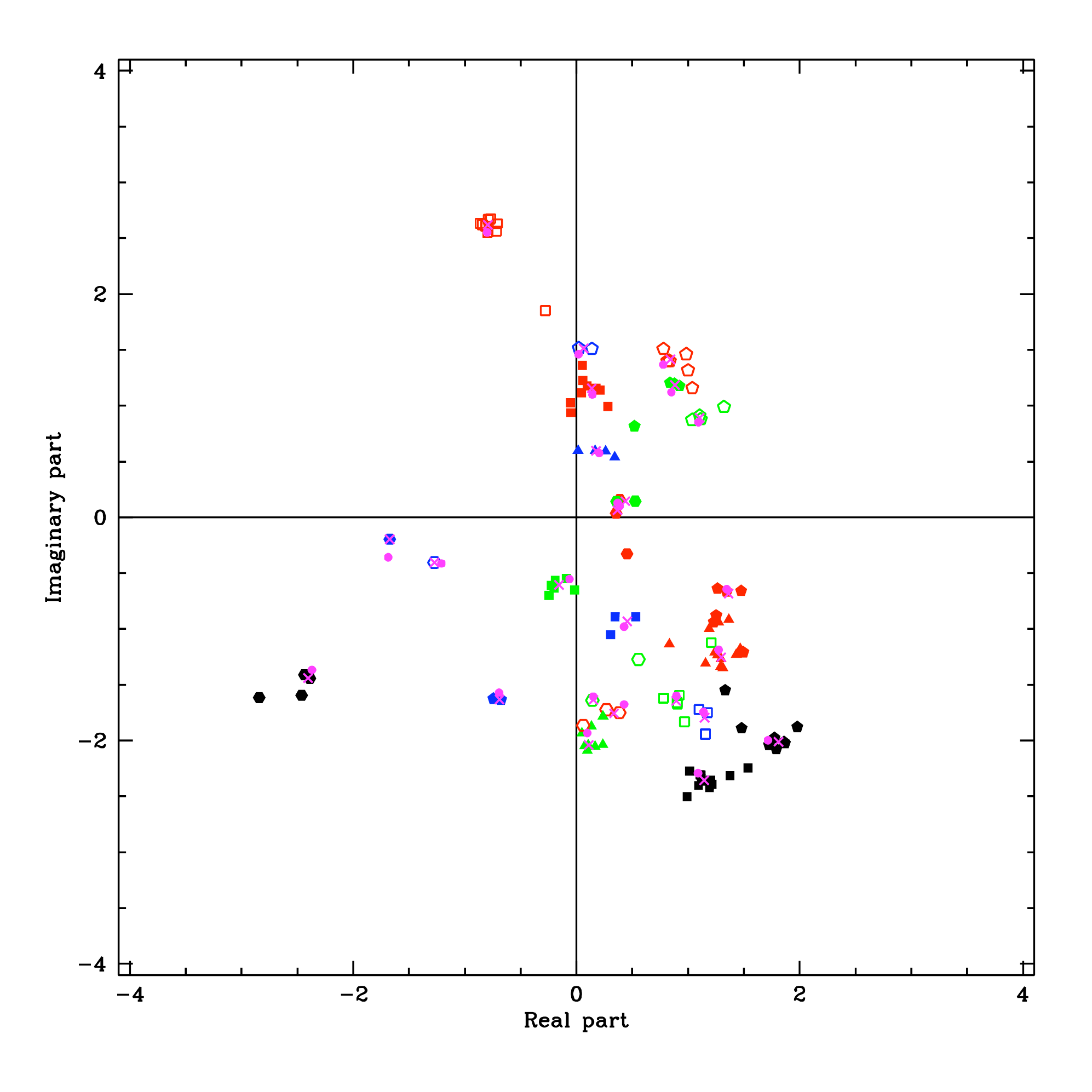}
\caption{Same as Figure \ref{originalvisibilities}, except the visibilities have been calibrated.  The simulated sky visibilities are again given by the magenta dots, while the best estimates from the calibration algorithm for these visibilities are given by the magenta x's.}
\label{adjustedvisibilities}
\end{figure}

As mentioned above, in addition to the antenna calibration parameters $\eta$ and $\varphi$, redundant calibration algorithms yield estimates $y_{ij}$ for the true sky visibilities.  In Figures \ref{legend}, \ref{originalvisibilities}, and \ref{adjustedvisibilities}, we show the visibility results that fall out of the calibration.  Figure \ref{legend} serves as a legend for the $uv$ plane, with different colors and symbols signifying the different \emph{unique} baselines found in a 4 by 4 square array.  These symbols are used to plot the uncalibrated input correlations measured by the interferometer (\emph{i.e.} the $c_{ij}$'s) on a complex plane in Figure \ref{originalvisibilities}.  Each color and symbol combination may appear multiple times in Figure \ref{originalvisibilities}, because two baselines that are identical in principle may give different readings in practice because of instrumental noise and the fact that different antennas will have different calibration parameters.  Indeed, one can see from Figure \ref{originalvisibilities} that the uncalibrated correlations are rather spread out over the complex plane, and do not seem close to the simulated true sky visibilities, which are denoted in the figure by solid magenta dots.

After calibration, however, one has estimates for the antenna calibration parameters, and thus these uncalibrated correlations can be corrected to yield measurements that are close to the correct sky visibilities.  One simply divides out the complex gain factors, since the measured correlations $c_{ij}$ are related to the true sky visibilities $y_{i-j}$ by $c_{ij} = g_i^* g_j y_{i-j}$.  The results are shown in Figure \ref{adjustedvisibilities}, where it is clear from the color and symbol combinations that identical baselines now yield measurements that cluster around the simulated true values, which are still given by the solid magenta dots.  The scatter that remains within a given set of identical baselines is due to instrumental noise (set at a level corresponding to an SNR of 10 for this simulation).  It should be noted, however, that while this is a perfectly workable method for computing estimates of the sky visibilities, it is unnecessary.  Redundant calibration outputs a parameter vector estimate $\mathbf{\hat{x}}$ (see Equation \ref{least^2}) that contains the true visibilities.  These are shown in Figure \ref{adjustedvisibilities} using magenta x's, and emerge from the calibration at no additional computational cost\footnote{In fact, it is not even strictly valid to consider any ``extra" computational cost associated with solving for the $y_{ij}$'s, for as one can see from Equation \ref{matrix1}, the inclusion of the $y_{ij}$'s is \emph{necessary} for redundant calibration to work.}.


\begin{figure}
\centering
\includegraphics[width=0.5\textwidth,trim=1.0cm 5.0cm 1.0cm 4.0cm, clip]{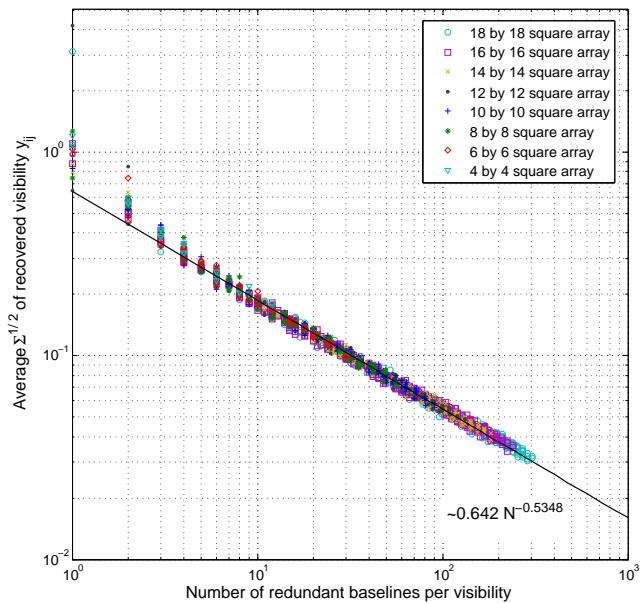}
\caption{Expected error bars on the estimated visibilities, as a function of the number of redundant baselines that go into determining each visibility.  A signal-to-noise (SNR) ratio of 1 is assumed, although in a realistic situation one would expect a more favorable SNR.}
\label{yerrors}
\end{figure}

Since the true visibilities (like the antenna gain parameters) are simply components of the parameter vector $\mathbf{x}$, we can again use the machinery of Equation \ref{rescaled error bars} to compute the expected error bars in the estimated visibilities.  The results are similar to what was found for the antenna calibration parameters, and are summarized in Figure \ref{yerrors}.  The errors are again found to asymptote to a $1/\sqrt{N}$ dependence, but here $N$ refers to the \emph{number of redundant baselines that go into determining a given visibility}, and not the total number of antennas in the array.  That the error bars do not depend on the total number of antennas is clear from Figure \ref{yerrors}, where data points from simulations of arrays of various sizes lie on top of each other.  It is also a property that is intuitively expected, since adding extra antennas increases the mathematical constraints on a given visibility only if the extra antennas form new baselines that correspond to that visibility's location on the $uv$ plane.

\subsection{Problems with the logarithmic implementation}
\label{logprobs}
While our simulations and real-world applications like those detailed in \citet{oldredund1,oldredund2,oldredund3,oldredund4,oldredund5,oldredund6} have shown above that taking logarithms of Equation \ref{toomanyvars} yields a linear system that can be successfully used to fit for calibration parameters and visibilities, this logarithmic implementation is not without its drawbacks.  In particular, the implementation has two undesirable properties:
\subsubsection{Phase Wrapping}
For the logarithmic implementation to work, the phase calibration parameter $\varphi$ needs to be close to zero for all antennas.  This is because one must take the complex logarithm of $c_{ij}=g_i^* g_j y_{i-j}$, which is an operation that is determined only up to additions or subtractions of multiples of $2 \pi$ in the imaginary part (\emph{i.e.} in the phase of the original number).  Thus. while the calibration solution may correctly recover the \emph{measured} correlations (since one is insensitive to additions or subtractions of $2\pi$ in the phase when re-exponentiating to find $c_{ij}=g_i^* g_j y_{i-j}$), it does not give correct phases for $g_i^*$, $g_j$, and $y_{i-j}$, which are ultimately the quantities of interest.  This is a problem even when only a small number of antennas have large phase calibration parameters, since to find the visibilities and antenna gain parameters redundant algorithms essentially performs a \emph{global} fit, where all antennas are coupled to each other via the visibilities.

\begin{figure}
\centering
\includegraphics[width=0.5\textwidth,trim=1.0cm 5.0cm 1.5cm 5.0cm, clip]{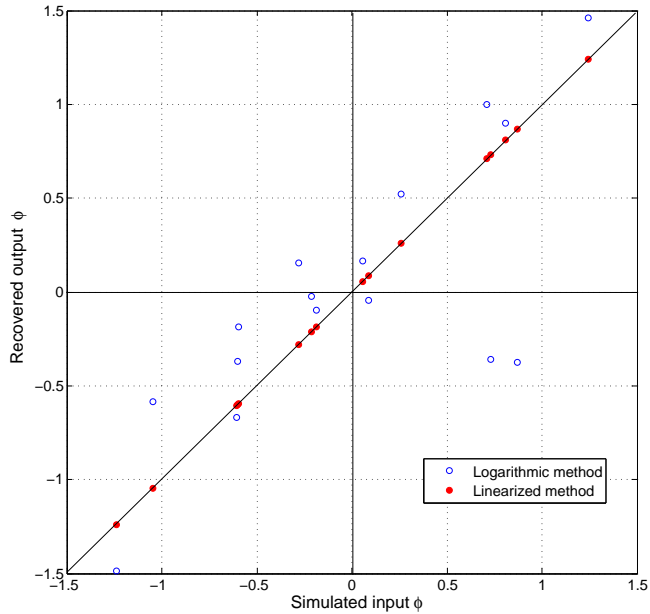}
\caption{Scatter plots of simulated vs. recovered antenna phase parameter $\varphi$ for a \emph{noiseless} 4 by 4 square array with relatively large phases.  The results from the logarithmic implementation described in Section \ref{log} are shown using open blue circles, whereas the results from the linear implementation described in Section \ref{linear} are shown using solid red circles.  Large phases are seen to affect the accuracy of the logarithmic method, but not the linear method.}
\label{2pifixed}
\end{figure}

In Figure \ref{2pifixed} we show the phase calibration results for a \emph{noiseless} simulation of a 4 by 4 array with phases that are significantly greater than zero.  The logarithmic method is shown using the open blue circles, and since a noiseless simulation should yield perfect parameter recovery, the scatter in the trend suggests a problem with the method.  A detailed examination of the simulation's numerical results reveals that it is indeed the multi-valued property of the complex logarithm that causes the trouble, and from the plot one also sees that a small number of very badly fit antennas can give rise to inaccuracies for all antennas, as is expected from the global nature of the fit.

It is important to note that even when the phase calibration fails because of large phases, the \emph{amplitude} calibration can still be implemented successfully.  This is because the taking of the complex logarithm means that in our system of equations, the amplitudes and phases separate into the real and imaginary parts respectively, as we can see from Equations \ref{logreal} and \ref{logim}.  The two calibrations are therefore completely decoupled, and instabilities in one do not affect the other.  The fact that an accurate gain calibration can be extracted regardless of the quality of one's phase calibration will be important in Section \ref{linearprobs}.

\subsubsection{Bias}
The logarithmic method is not unbiased, in the sense that ensemble averages of noisy simulations do not converge to the true simulated parameter values.  While Equation \ref{least^2} can be shown to be unbiased \citep{MaxMaps}, the proof assumes that the noise covariance matrix $\mathbf{N}$ is Gaussian and that the weights used in the fits (encoded by $\mathbf{N}^{-1}$) are independent of the data.  In our case both assumptions are violated.  While we may assume that $n_x$ and $n_y$ in Equation \ref{nxny} are Gaussian distributed, this is not the case in amplitude/phase angle space.  In addition, since the diagonal entries of $\mathbf{N}$ are taken to be $1/|c|$, the measured data enter the least squares fit (Equation \ref{least^2}) not just through the data vector $\vec{d}$ but also via $\mathbf{N}^{-1}$, making the fitting process a nonlinear one.

\begin{figure}
\centering
\includegraphics[width=0.5\textwidth,trim=1.0cm 5.0cm 1.5cm 5.0cm, clip]{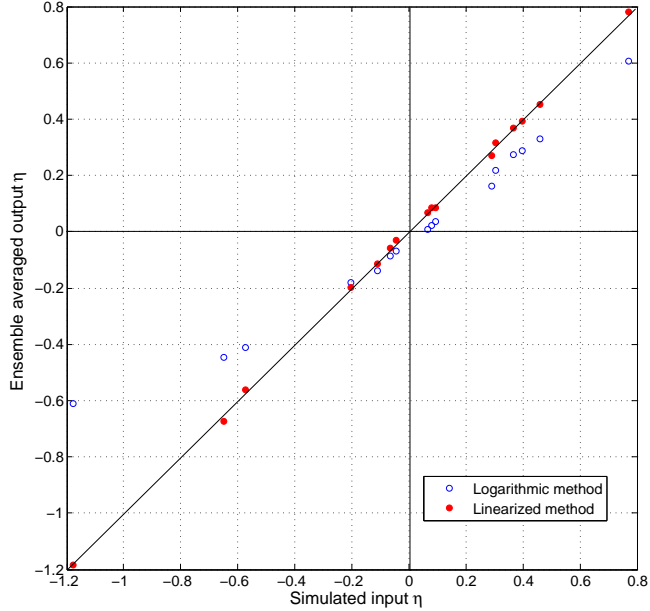}
\caption{Scatter plots of simulated vs. ensemble averaged recovered antenna phase parameter $\varphi$ for a 4 by 4 square array with a signal-to-noise ratio of 2.  The results from the logarithmic implementation described in Section \ref{log} are shown using open blue circles, whereas the results from the linear implementation described in Section \ref{linear} are shown using solid red circles.  The logarithmic algorithm gives a biased result, whereas the linear implementation does not.}
\label{Biasplot}
\end{figure}

The bias in the method can be easily seen in Figure \ref{Biasplot}, where the blue dots show the simulated results for $\eta$, averaged over 90 random ensemble realizations of a 4 by 4 array with an SNR of 2.  There is clearly a systematic bias in the recovered output $\eta$'s.

\subsection{The Linearized Approach}
\label{linear}
Since both of the problems discussed in the previous section arise at least partly from taking the logarithm of Equation \ref{prelog}, we now propose a method that does not require that step.  As an alternative to taking logarithms, one can instead linearize the equations in Section \ref{our}.  We do so by Taylor expanding our expressions about some fiducial guesses $\eta_0$, $\varphi_0$, and $y_0$ for the parameters $\eta$, $\varphi$, and $y_0$.  The gain of a particular antenna, for example, becomes
\begin{subequations}
\begin{eqnarray}
g & \equiv & e^{\eta + i \varphi} \\
&=& g_0 + \frac{\partial g}{\partial \eta}  \bigg|_{\eta_0, \varphi_0} \!\!\!\!\!\!  (\eta - \eta_0) +  \frac{\partial g}{\partial \varphi}  \bigg|_{\eta_0, \varphi_0} \!\!\! \!\!\!  (\varphi - \varphi_0) + \dots \\
&=& \exp \left( \eta_0 + i \varphi_0 \right) \left[ 1+ (\eta - \eta_0) + i (\varphi - \varphi_0 ) \right] + \dots  \qquad
\end{eqnarray}
\end{subequations}
Along with our fiducial guess for the true correlations (the $y$'s), we define guesses for our measured correlations:
\begin{equation}
 c_{ij}  ^0 \equiv y_{i-j}^0 \exp \left( \eta_i^0 - i \varphi_i^0 \right) \exp \left( \eta_j^0 + i \varphi_j^0 \right).
\end{equation}
In addition, we define the \emph{deviation} $\delta_{ij}$ of the guessed correlations from the measured correlations
\begin{equation}
\delta_{ij} \equiv  c_{ij}  -  c_{ij}  ^ 0
\end{equation}
as well as an analogous quantity for the true sky correlations
\begin{equation}
y_{i-j}^1 \equiv y_{i-j} - y_{i-j}^0.
\end{equation}
To calibrate our telescope, we take the measured deviations $\delta_{ij}$ as our inputs and compute the corrections to the guessed antenna parameters and guessed true sky correlations that are required to match these deviations.  Plugging our definitions and expansions into $c_{ij} = g_i^{*} g_j y_{i-j}$ gives
\begin{equation}
\delta_{ij} \approx \exp \left( \eta_i^0 - i \varphi_i^0 \right) \exp \left(\eta_j^0 + i \varphi_j ^ 0 \right) 
\left[ y_{i-j}^1 + y_{i-j}^0 \left( \Delta \eta_i + \Delta \eta_j - i \Delta \varphi_i + i \Delta \varphi_j \right) \right]
\end{equation}
where we have defined $\Delta \eta \equiv \eta - \eta_0 $, $\Delta \varphi \equiv \varphi - \varphi_0$, and have discarded second order terms in the small quantities $\Delta \eta$, $\Delta \varphi$, and $y^1$.  The result is a linear system in $\Delta \eta$, $ \Delta \varphi$, and $y^1$, which means we can use the matrix formalism presented in the previous section, the only differences being that the $\eta$ and $\varphi$ equations are now coupled, and that the matrix analogous to the $\mathbf{A}$ used in the logarithmic method --- which we will call $\mathbf{B}$ for the linearized method --- now depends on the fiducial guesses as well as the array layout.  A least-squares fit can once again be employed to solve for the $\Delta \eta$'s, the $\Delta \varphi$'s, and $y_{i-j}^1$.  Once the fit has been obtained, it can be used to update our fiducial guesses, and if desired, one can continue to improve the quality of the calibration by repeating the algorithm iteratively --- the updated calibration parameters are simply used as the fiducial guesses for the next cycle of fitting.  Note that unlike before, it is unnecessary to use a weighted fit \emph{i.e.} one can set $\mathbf{N=I}$ in Equation \ref{least^2}.  This is because the linearized algorithm works with visibilities directly (as opposed to, say, their logarithm or some other function of the visibilities), and since our assumption of uncorrelated baseline errors means that individual baselines have the same noise level, all correlations should be weighted the same in the fitting.

\begin{figure*}
\centering
\includegraphics[width=1.0\textwidth]{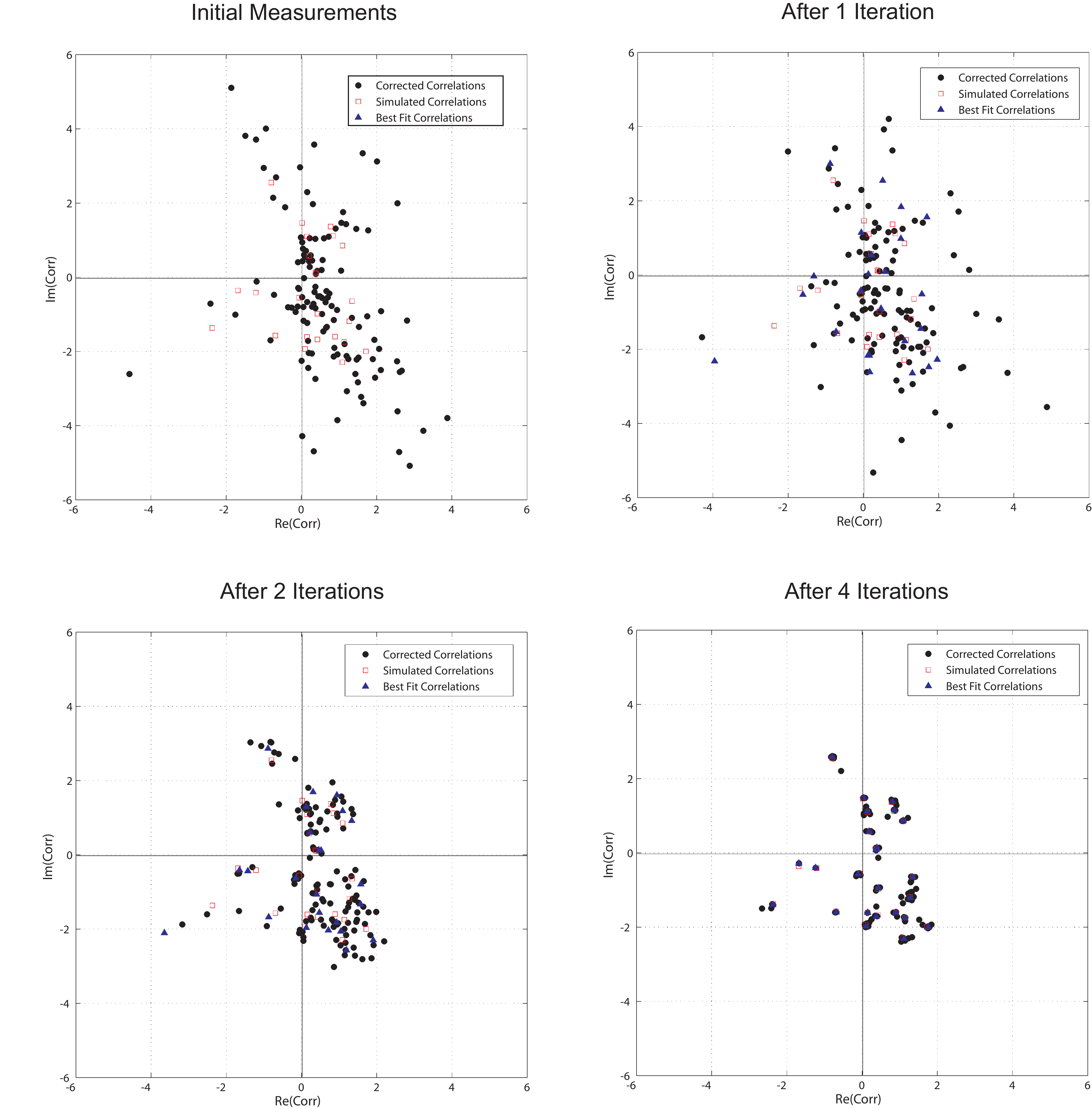}
\caption{Visibility simulations, measurements, and estimated visibilities plotted on complex planes.  The top left plane shows the simulated and measured visibilities before any calibration scheme has been applied.  The top right plane shows the simulated visibilities, corrected visibilities, and best fit visibilities after one iteration.  The bottom planes show the same quantities after two and four iterations respectively.}
\label{iterations}
\end{figure*}

In Figure \ref{iterations}, we show plots of simulated and recovered visibilities on the complex plane as one steps through each iteration of the algorithm.  The top left plane shows the simulated correlations as well as the measured correlations.  After one iteration, we plot the corrected visibilities as well as the estimated true visibilities in the top right.  The recovered visibilities are seen to be still quite different from the simulated ones.  The bottom left and bottom right planes show the second and fourth iterations respectively, and one can see that the recovered visibilities converge toward their simulated values.

Scatter plots showing the correlation between the simulated and estimated antenna calibration parameters are qualitatively identical to Figures \ref{logeta} and \ref{logphi} for the logarithmic implementation, so we do not show them here.   However, the linearized approach solves the two problems with the logarithmic algorithm discussed in Section \ref{logprobs}.  In Figure \ref{2pifixed}, the filled red circles representing the linearized method show a perfect recovery of simulated phases in the noiseless limit.  This demonstrates that the linearized algorithm can tolerate arbitrarily large phases, which the logarithmic method could not.  Figure \ref{Biasplot} shows that the ensemble-averaged parameter estimates coming out of the linearized method converge to their true values, demonstrating that the linearized algorithm is unbiased.  That this is the case is particularly important if one applies the algorithm to highly noisy systems, for then one can compensate for the low SNR by using massive arrays with large numbers of baselines and by averaging the results over long time periods\footnote{That is, provided any natural drift in the calibration parameters occur over longer timescales than the required averaging time.}.

\subsection{Numerical issues for the linearized approach}
\label{linearprobs}
Although the linearized algorithm evades the problems with the logarithmic approach, it is slightly more complicated to put into practice.  Below we discuss two problems that may arise in the numerical implementation of the method and how to solve them.

\subsubsection{Convergence}
Bad fiducial guesses can result in misfits, although such fits can be easily identified and improved upon.  Given that our basic measurement equation (Equation \ref{toomanyvars}) is nonlinear in our parameters, the linearization that we have presented in this section can only be expected to stably yield good fits when our fiducial guesses are reasonably close to their true values.  In a naive implementation of the algorithm, bad fiducial guesses can result in an inability to iterate out of a local maximum in goodness-of-fit.  However, bad fits can be readily identified by simply computing the goodness-of-fit parameter, and if necessary, one can repeat the fit with a better search algorithm such as simulated annealing to find the correct maximum.

Another way to avoid misfits is to use the results of the logarithmic algorithm as our initial guess for the linearized method.  Even though the phase calibration has been shown to be unreliable when the phases are large, the amplitude calibration can still be trusted since it comes from solving a completely separate system of equations.  The fact that the amplitudes estimated using the logarithmic method are biased is not an issue, since our fiducial guess serves merely as a starting point for our iterative linearized scheme.

An attractive approach to implementing our method in practice could be to update the calibration on the timescale over which the calibration parameters are expected to drift by non-negligible amounts (say once per second or once per minute), using the previously determined calibration parameters as the initial guess and performing merely a single iteration each time. Avoiding multiple iterations in this fashion has the advantage that the noise from any given observation affects the output only linearly, thus avoiding nonlinear mappings that can potentially bias the results.
The optimal duration between calibrations should be determined to minize the expected errors: if we calibrate too frequently, the calibration parameters get unnecessarily noisy because so little data is used to determine them, while if we calibrate too rarely, the true calibration parameters can drift appreciably before we recalibrate.
An alternative way to determine this tradeoff is to re-estimate the calibration parameters more frequently than needed and then smooth the time series, 
each time replacing the calibration parameter vector actually used 
by $p$ times the new estimate and $(1-p)$ times the last estimate, and determining which choice for $p\in(0,1)$ minimizes the rms calibration errors.

\subsubsection{Computational Cost}
The linear method is more computationally intensive than the logarithmic one, especially if more than one iteration is required.  However, the computational cost is still acceptable if one takes advantage of the sparseness of $\mathbf{B}$.  At first sight, the setting of $\mathbf{N=I}$ in Equation \ref{least^2} for the linearized method but not for the logarithmic method would seem to make the linear method computationally quicker.  However, the presence of a non-trivial $\mathbf{N}$ in the logarithmic method does not increase the computational cost, for in the absence of cross-talk $\mathbf{N}$ is diagonal.  The computational cost of its inversion is therefore negligible, and its presence in Equation \ref{least^2} simply changes the weighting of matrix elements when finding matrix products.

That the linear method is slightly more computationally intensive than the logarithmic method is due to the fact that the matrices for the linear method are larger than those in the logarithmic method.  This is because the amplitude and phase calibrations do not decouple in the linearized scheme, and all the parameters must be solved for in one big system.  In particular, all vectors are now twice as long as before, since a system where gains and phases are coupled is mathematically equivalent to one where the real and imaginary parts of all complex numbers are dealt with together.  This has its greatest computational impact on the matrix inversion of $\mathbf{B^t B}$, which can scale as strongly as the vector length cubed if one simply implements the most straightforward matrix inversion routines.  Since this inversion must be performed every time the fiducial values (which are part of $\mathbf{B}$) are updated, the computational time also scales linearly with the number of iterations, and so naively the linearized method seems rather expensive.  However, the matrix $\mathbf{B}$ is by construction always sparse, which means the matrix inversion can be performed more efficiently.  As an example, one may use the conjugate gradient method, where the sparseness of $\mathbf{B}$ means that the inversion scales linearly with vector length, improving the numerical cost scaling from $\cal{O}$$(N^3)$ to $\cal{O} $$(N)$.  This is very similar to the mapmaking approach successfully implemented by the WMAP team in  \citet{WMAP}.

It should also be noted that if only one iteration is required and the gain parameters drift only by small amounts over time, then the linearized method can be extremely efficient over long integrations.  This is because a \emph{series} of calibrations can then be performed using the same fiducial values for the antenna parameters, and thus the matrix $\mathbf{[B^t B]^{-1} B^t}$ never changes and can be computed once and for all\footnote{In principle, one also requires that transients are not present in the data, since transients may cause sudden changes in the visibilities $y_{ij}$ and thus one cannot assume that the same fiducial guesses are good ones over all time.  In practice, however, transients can be easily identified in the data, and one can simply ignore the data taken over time intervals where transients dominate.}.  One then simply stores this matrix, which is applied to the data whenever calibration is required.  Note that this cannot be done with the logarithmic approach, since the relevant matrix there is $\mathbf{[A^t N^{-1} A]^{-1} A^t N^{-1}}$ and from Equation \ref{Ncovar} we see that $\mathbf{N}$ changes as the measured correlations change, so the combination $\mathbf{[A^t N^{-1} A]^{-1} A^t N^{-1}}$ must be recomputed every time one would like to calibrate.

If an instrument with $N$ antennas possesses a very large number of redundant baselines (like any reasonably large omniscope, say, where almost all of the $\sim N^2/2$ baselines are redundant), 
Figure \ref{yerrors} suggests that the errors will be extremely small, especially if the signal-to-noise ratio is high. 
With such an instrument, one will be able to calibrate accurately by utilizing merely a small subset of the redundant baselines; for example, if each antenna is correlated with merely 20 others, 
the resulting system of equations will still be safely overdetermined while keeping the size of the $\mathbf{B}$-matrix and the computational cost reasonable.
This is important for any array utilizing fast Fourier transforms for the correlation process, since its attractive $N\log_2 N$ cost scaling would be lost if an $\cal{O}$$(N^2)$ correlator were needed for calibration.

In summary, our proposed calibration method can be performed rather cheaply, especially when one compares the computational cost to that of initially computing the interferometric correlations.  To see this, note that one must compute correlations roughly once every $10^{-8}\,\textrm{seconds}$ for an interferometer operating at a frequency of $\sim 100\,\textrm{MHz}$.  This is much more expensive than the calibration step (especially if one incorporates the computational tricks suggested in this section), which only needs to be performed, say, once every second or minute.
In other words, the correlation needs to be repeated on the timescale on which the electromagnetic waves oscillate whereas the calibration only needs to be repeated the time scale on which the calibration parameters drift.

\section{A Generalized Calibration Formalism: Calibration as a Perturbative Expansion}
\label{generalized}
In previous sections, the calibration algorithms that were discussed represented two idealized limits: with the bright point source calibration (Section \ref{traditional}), the existence of an \emph{isolated} bright point source was key, while with redundant baseline algorithms (Section \ref{our}) a large number of baselines of equal length and orientation were required.  Both algorithms were ``zeroth order algorithms" in that they required exact adherence to these requirements.  In this section, we perturb around these idealized assumptions, and show that point source calibration and redundant calibration are simply two extremes in a generalized continuum of calibration schemes.

Consider the sky response (\emph{i.e.} the visibility) from a baseline $\mathbf{b}$ formed by two antennas at locations $\mathbf{r}_i$ and $\mathbf{r}_j$ (\emph{i.e.} $\mathbf{b} \equiv \mathbf{r}_i - \mathbf{r}_j$).  Assuming that our array is coplanar, we have
\begin{equation}
\label{basiceqn}
c(\mathbf{b}) = \int_{sky} g_i^* g_j \frac{J (\boldsymbol{\theta}) B(\boldsymbol{\theta})}{\sqrt{1-\theta_x^2 - \theta_y^2}} \exp \left( -\frac{i 2 \pi \mathbf{b} \cdot \boldsymbol{\theta}}{\lambda} \right) d^2 \boldsymbol{\theta},
\end{equation}
where $J(\boldsymbol{\theta})$ is the intensity of the sky signal from the $\boldsymbol{\theta}$ direction, and $B(\boldsymbol{\theta})$ is the primary beam.  Suppose we now define an effective sky signal $I ( \boldsymbol{\theta}) \equiv 
\frac{J (\boldsymbol{\theta}) B(\boldsymbol{\theta})}{\sqrt{1-\theta_x^2 - \theta_y^2}} $.  Writing this in terms of its Fourier space description
\begin{equation}
\label{deffourier}
I (\boldsymbol{\theta}) = \int \tilde{I} ( \mathbf{k} ) \exp ( i 2 \pi \mathbf{k} \cdot \boldsymbol{\theta} ) d^2 \mathbf{k}
\end{equation}
and inserting this into Equation \ref{basiceqn}, we get
\begin{equation}
\label{fourier}
c( \mathbf{b} ) = g_i^* g_j \tilde{I} \left( \frac{\mathbf{b}}{\lambda} \right),
\end{equation}
which is a well-known result: baselines of an interferometer probe different Fourier modes of the effective sky signal.

Now suppose that our interferometer possesses a number of nearly-redundant baselines.  In other words, suppose that there exist a set of vectors $\mathbf{b}_0^{\alpha}=\{ \mathbf{b}_0^{1},\mathbf{b}_0^{2},\dots \}$ on the $uv$ plane that baselines tend to cluster around.  Nearly-redundant baselines are considered part of the same cluster, and are associated with a single $\mathbf{b}_0^{\alpha}$.  On the other hand, any baselines that are isolated on the $uv$ plane are associated with ``their own" $\mathbf{b}_0^{\alpha}$.  Thus, for an array with $N$ antennas the index $\alpha$ runs from $1$ to $N(N-1)/2$ if there are zero near or exact redundancies in the baselines, and to some number substantially smaller than $N(N-1)/2$ if there are many near or exact redundancies.  In other words, we have grouped the baselines into nearly-redundant clusters.

We now re-parameterize our baseline distribution in terms of $\delta \mathbf{b}_{ij} \equiv \mathbf{b}_{ij} - \mathbf{b}_{0}^{\alpha}$, which measures each baseline's deviation from perfect redundancy.  Since $\delta \mathbf{b}_{ij}$ is a small quantity, we can Taylor expand Equation \ref{fourier}:
\begin{subequations}
\begin{eqnarray}
\label{taylor1}
c_{ij} &=& g_i^* g_j \tilde{I} \left( \frac{\mathbf{b}_0^{\alpha}+\delta \mathbf{b}_{ij}}{\lambda} \right) \\
\label{taylor2}
& \approx & g_i^* g_j \left[ \tilde{I}\left( \frac{\mathbf{b}_0^{\alpha}}{\lambda} \right)+ \mathbf{\nabla}_{\mathbf{u}}\tilde{I} \bigg|_{\mathbf{b}=\mathbf{b}_0^{\alpha}}\cdot \frac{\delta \mathbf{b}_{ij}}{\lambda} + \dots \right], \qquad
\end{eqnarray}
\end{subequations}
where $\mathbf{\nabla}_{\mathbf{u}}$ denotes a two-dimensional gradient in the $uv$ plane.  If we define $c_0^{\alpha} \equiv  \tilde{I}\left( \frac{\mathbf{b}_0^{\alpha}}{\lambda} \right)$, we can rewrite our equation as
\begin{equation}
\label{linearfirstorder}
c_{ij} \approx g_i^* g_j c_0^{\alpha} \left( 1 + \mathbf{h}_0^{\alpha} \cdot \frac{\delta \mathbf{b}_{ij}}{\lambda} + \dots \right),
\end{equation}
where $\mathbf{h}_0^{\alpha} \equiv \mathbf{\nabla} \ln \tilde{I}  \bigg|_{\mathbf{b}=\mathbf{b}_0^{\alpha}}$.  

For algebraic brevity, we proceed using the logarithmic method of Section \ref{log}, the generalization to the linearized method being straightforward.  Taking logarithms of both sides yields
\begin{eqnarray}
\label{firstexp}
\ln c_{ij} &\approx& (\eta_i + \eta_j) + i (\varphi_j - \varphi_i ) + \ln c_0^{\alpha} + \ln \left( 1 + \mathbf{h}_0^{\alpha} \cdot \frac{\delta \mathbf{b}_{ij}}{\lambda} \right) \nonumber \\
&\approx& (\eta_i + \eta_j) + i (\varphi_j - \varphi_i )  + \ln c_0^{\alpha} + \mathbf{h}_0^{\alpha} \cdot \frac{\delta \mathbf{b}_{ij}}{\lambda},
\end{eqnarray}
where we have assumed that $\mathbf{h}_0^{\alpha} \cdot \frac{\delta \mathbf{b}_{ij}}{\lambda}$ is small in order to expand the last logarithm.

Before we proceed, it is worth examining the precise conditions under which our Taylor series expansions are valid.  The crucial quantity is $\mathbf{h}_0^{\alpha} \cdot \frac{\delta \mathbf{b}_{ij}}{\lambda}$, which will be small if either $\mathbf{h}_0^{\alpha} $ or $\frac{\delta \mathbf{b}_{ij}}{\lambda}$ are small.  Plugging Equation \ref{deffourier} into our definition of $\mathbf{h}$, we have
\begin{equation}
\mathbf{h} = \frac{\mathbf{\nabla} \tilde{I}}{\tilde{I}} = - i 2 \pi \frac{\int \boldsymbol{\theta} I ( \boldsymbol{\theta} ) \exp ( -i 2 \pi \mathbf{u} \cdot \boldsymbol{\theta} ) d^2 \boldsymbol{\theta}}{\int I ( \boldsymbol{\theta} ) \exp ( -i 2 \pi \mathbf{u} \cdot \boldsymbol{\theta} ) d^2 \boldsymbol{\theta}},
\end{equation}
which is a quantity whose magnitude is at most the size of the primary beam in radians.  This dependence on the primary beam means that $\mathbf{h}_0^{\alpha} \cdot \frac{\delta \mathbf{b}_{ij}}{\lambda}$ can be made to be small in different ways depending on the instrument one is considering:
\begin{enumerate}
\item For widefield instruments, the primary beam width is on the order of $1\,\textrm{radian}$, so one requires $| \delta \mathbf{b}_{ij} | \ll \lambda$.
\item For instruments with narrow primary beams such as the VLA, we have $| \mathbf{h}| \ll 1$, so the deviations $\delta \mathbf{b}_{ij}$ from perfect redundancies need not be small compared to the wavelength.
\end{enumerate}

If the conditions listed above are satisfied, then calibrating our radio telescope is similar to the zeroth-order case, in the sense that it is once again tantamount to solving Equation \ref{firstexp}.  Here, however, we are \emph{not} guaranteed to have enough equations to solve for all the relevant variables.  In addition to the $2N$ antenna calibration parameters ($\eta$'s and $\varphi$'s), we now potentially have up to $3N(N-1)$ numbers to solve for --- the $\ln c_0^{\alpha}$ term contributes up to $N(N-1)$ numbers, since as we discussed above $\alpha$ can run to $N(N-1)/2$ in a worst-case scenario, and specifying a single $\ln c_0^{\alpha}$ requires two numbers: a real part and an imaginary part; the $\mathbf{h}_0^{\alpha}$ term contributes up to $2N(N-1)$ numbers --- twice that required to specify the $\ln c_0^{\alpha}$ terms, because $\mathbf{h}_0^{\alpha}$ is the logarithmic gradient of $c_0^{\alpha}$ in the $uv$ plane, which is two dimensional.  The $\delta \mathbf{b}_{ij}$'s do not need to be solved for since they can be computed from the layout of the array.  Putting this all together, we see that in general there can be up to $3N(N-1)$ unknowns.

To properly solve Equation \ref{firstexp}, one must therefore find ways to reduce the range of the index $\alpha$ so that the number of equations exceeds the number of unknowns.  In the next few subsections, we will consider different scenarios in which this is the case.

\subsection{Zeroth Order: Perfect Point Sources or Perfect Redundancies}
\label{zero}
The zeroth order solution to Equation \ref{firstexp} involves setting up situations where the $\mathbf{h}_0^{\alpha} \cdot \frac{\delta \mathbf{b}_{ij}}{\lambda}$ term can be set to zero.  There are two ways to accomplish this:
\begin{enumerate}
\item \textbf{Design an array with perfect redundancies, so that $\delta \mathbf{b}_{ij} = 0$}.  The form of $\mathbf{h}_0^{\alpha}$ is then irrelevant (which is another way of saying that we can calibrate with any sky signal), and the only requirement is for the number of \emph{unique} baselines to be small enough that the maximum possible $\alpha$ is modest and there are only a small number of $c_0^{\alpha}$ terms to fit for.  This is precisely the redundant calibration algorithms discussed in Sections \ref{our} to \ref{linearprobs}.
\item \textbf{Have some knowledge of the sky signal used for calibration, so that $c_0^{\alpha}$ and $\mathbf{h}_0^{\alpha}$ are known.}  For example, if one uses a phased array to image a point calibrator source, then $c_0^{\alpha}$ becomes a real constant independent of $\alpha$, while the $h_0^{\alpha}$'s all vanish.  This is the point source calibration discussed in Section \ref{traditional}.
\end{enumerate}

\subsection{First Order: Unknown Sources and Near Perfect Redundancies}
If one is unable to satisfy the conditions listed in Section \ref{zero}, it may still be possible to calibrate by solving for $c_0^{\alpha}$ and $\mathbf{h}_0^{\alpha}$ as well as all the calibration parameters simultaneously.  For example, taking the real part of Equation \ref{firstexp} gives
\begin{equation}
\label{firstorderbaselinecorrection}
\ln |c_{ij}| = (\eta_i + \eta_j)  + \ln |c_0^{\alpha}| + \textrm{Re}(h_{0u}^{\alpha}) \frac{\delta {b}_{ij}^u}{\lambda}+\textrm{Re}(h_{0v}^{\alpha}) \frac{\delta {b}_{ij}^v}{\lambda},
\end{equation}
which can be written as a matrix system:
\begin{equation}
\left(\begin{array}{c}
\ln \left| c_{12} \right| \\
\ln \left| c_{23} \right| \\
\ln \left| c_{34} \right| \\
\ln \left| c_{45} \right| \\
\ln \left| c_{56} \right| \\
\vdots \\
\ln \left| c_{29} \right| \\
\ln \left| c_{19} \right|
\end{array} \right)\\
=
\left( \begin{array}{cccccccccccc}
1 & 1 & \dots & 1 & 0 & \dots & \frac{\delta {b}_{12}^u}{\lambda} & 0 & \dots & \frac{\delta {b}_{12}^u}{\lambda} & 0 & \dots \\
0 & 1 & \dots & 1 & 0 & \dots & \frac{\delta {b}_{23}^u}{\lambda} & 0 & \dots & \frac{\delta {b}_{23}^u}{\lambda} & 0 & \dots \\
0 & 0 & \dots & 1 & 0 & \dots & \frac{\delta {b}_{34}^u}{\lambda} & 0 & \dots & \frac{\delta {b}_{34}^u}{\lambda} & 0 & \dots \\
0 & 0 & \dots & 0 & 1 & \dots & 0 & \frac{\delta {b}_{45}^u}{\lambda} & \dots & 0 & \frac{\delta {b}_{45}^u}{\lambda} & \dots \\
0 & 0 & \dots & 0 & 1 & \dots & 0 & \frac{\delta {b}_{56}^u}{\lambda} & \dots & 0 & \frac{\delta {b}_{56}^u}{\lambda} & \dots \\
\vdots & \vdots &  & \vdots & \vdots &  & \vdots & \vdots &  & \vdots & \vdots &  \\
0 & 1 & \dots & 0 & 0 & \dots & 0 & 0 & \dots & 0 & 0 & \dots \\
1 & 0 & \dots & 0 & 0 & \dots & 0 & 0 & \dots & 0 & 0 & \dots \\
\end{array} \right) \\
\left(\begin{array}{c}
\eta_1 \\
\eta_2 \\
\vdots \\
\ln \left| c_0^1 \right| \\
\ln \left| c_0^2 \right| \\
\vdots \\
\textrm{Re}(h_{0u}^{1}) \\
\textrm{Re}(h_{0u}^{2})  \\
\vdots \\
\textrm{Re}(h_{0v}^{1}) \\
\textrm{Re}(h_{0v}^{2})  \\
\vdots \\
\end{array} \right),
\end{equation}
where for illustrative purposes we have chosen a nine-antenna system with the first three baselines almost-redundant.  This set of equations can again be solved using Equation \ref{least^2}, provided the design of the interferometer is such that the baselines can be grouped into a relatively small number of near-redundant clusters.  Fundamentally, the method here is the same as it has been in all previous sections --- we have simply written the interferometer response as a linear function of the calibration parameters and the sky signal, and have found ways to reduce the number of parameters (whether by mathematical approximation, interferometer design, or careful selection of calibration source) so that the equations are solvable.  In the first order expansion considered in this section, if $r$ denotes the number of $\mathbf{b}_0^{\alpha}$'s with two or more baselines clustered around them and $l$ denotes the number of completely isolated baselines, the calibration can be solved for provided our array satisfies
\begin{equation}
\label{varcounting}
\frac{N(N-1)}{2} > N + 3r +l,
\end{equation}
where as usual $N$ is the number of antenna elements in the array.  The left hand side is simply the number of measured complex correlations, while the right hand side counts the number of complex unknowns we need to solve for: $N$ complex antenna gains, $r$ complex visibilities at each cluster center, $2r$ complex numbers quantifying deviations in the complex visibilities due to departures from perfect redundancy in the two directions of the $uv$ plane, and $l$ complex visibilities for the isolated baselines.  For the $s \times s$ regular square arrays simulated in Section \ref{basic}, the condition is satisfied if $s \ge 4$.

Like before, however, one must be cognizant of possible degeneracies in the matrix system.  In particular, consider a cluster comprising of one or two baselines.  In fitting for this cluster's visibility, we have one or two inputs (\emph{i.e.} correlations) but need to solve for three parameters: $c_0^{\alpha}$, $h_{0u}^{\alpha}$, and $h_{0v}^{\alpha}$.  Our matrix system thus becomes singular.  The same problem arises if either $\delta b_{ij}^{u}$ or $\delta b_{ij}^{v}$ are identical for a large number of baselines in a particular cluster.  For instance, if all but one of the baselines in a cluster are perturbed in precisely the same way, then as far as the cluster goes we effectively only have two different baselines and cannot fit for three parameters.  These degeneracies can be dealt with in a similar fashion as before: one simply imposes extra constraints on the system of equations, like requiring that $h_{0u}^{\alpha}$ and $h_{0v}^{\alpha}$ be zero for single-baseline clusters.

\begin{figure}
\centering
\includegraphics[width=0.5\textwidth,trim=1.2cm 4.5cm 1.5cm 5.0cm, clip]{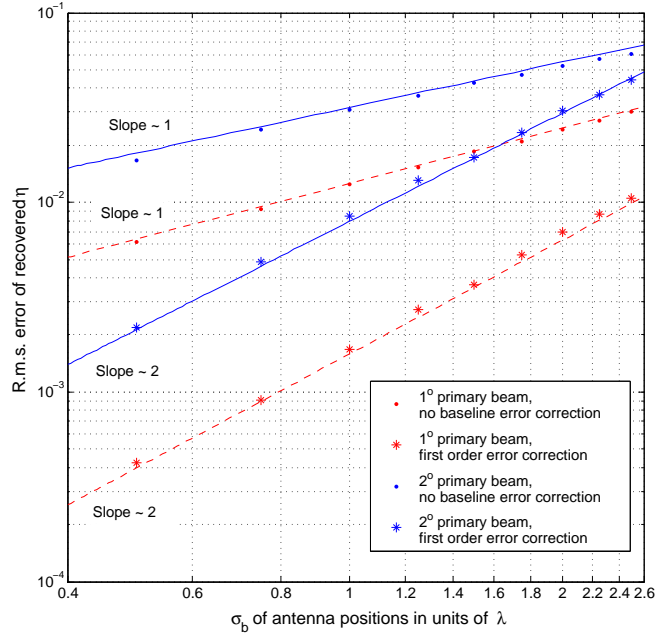}
\caption{R.m.s. errors in recovered $\eta$ for one antenna in a noiseless $4 \times 4$ square array as a function of antenna position spread $\sigma_b$.  Shown in red dashed lines are the results for a primary beam with width $1^{\circ}$, while the results for a primary beam with width $2^{\circ}$ are shown using the blue solid lines.  The solid circles denote results from an algorithm that does not correct for baseline position errors, and the stars denote results from an algorithm that corrects these errors to first order.}
\label{differentorders}
\end{figure}

Simulation results for a $4 \times 4$ square array with Gaussian antenna position errors are shown in Figure \ref{differentorders}, where we show the root mean square errors in the recovered gains $\Delta \eta$ as a function of the spread $\sigma_b$ in the Gaussian from which the position errors are drawn.  The simulations were run with precisely the same parameters as those in Section \ref{basic}, except with zero instrumental noise, so any non-zero error is due purely to position errors (\emph{i.e.} to the non-perfect redundancy of baselines).  Plotted using the solid circles are the results that one obtains if one simply ignores the fact that the baselines are not perfectly redundant \emph{i.e.} if one applies a zeroth order algorithm based on Equation \ref{prelog}.  The errors are clearly linear in $\sigma_b$.  Correcting for baseline errors to first order by implementing an algorithm based on Equation \ref{linearfirstorder} gives errors that are smaller and quadratic in $\sigma_b$.  It is also clear from the plot that the wider the primary beam of an instrument, the greater the errors in the calibration.  Formally, this is because the beam width is directly proportional to the size of the linear baseline error correction term, as we noted above.  This result can also be understood intuitively by considering the effect that a primary beam has on $uv$ visibilities.  A narrow primary beam acts as a wide convolution kernel on the $uv$ plane, which effectively ``averages" together visibilities over a wide neighborhood.  This smoothes the visibilities on the $uv$ plane, making the errors incurred by baseline errors more amenable to a perturbative correction.

In principle, there is no need to stop at the first order correction.  One could, for instance, correct baseline errors to second order by using the following equation as the basic measurement equation:
\begin{equation}
\label{quadorg}
c_{ij} \approx g_i^* g_j c_0^{\alpha} \left( 1 + \mathbf{h}_0^{\alpha} \cdot \frac{\delta \mathbf{b}_{ij}}{\lambda} + \frac{\delta \mathbf{b}_{ij}^t}{\lambda} \cdot \mathbf{j}_0^{\alpha} \cdot \frac{\delta \mathbf{b}_{ij}}{\lambda}\dots \right),
\end{equation}
where $\mathbf{h}_0^{\alpha} \equiv \mathbf{\nabla} \ln \tilde{I}  \bigg|_{\mathbf{b}=\mathbf{b}_0^{\alpha}}$ as before, and $\mathbf{j}_0^{\alpha}$ is a symmetric matrix of second derivatives of $\tilde{I}$ evaluated at $\mathbf{b}_0^{\alpha}$ and normalized by $\tilde{I}\left( \frac{\mathbf{b}_0^{\alpha}}{\lambda} \right)$.  It is important to note, however, the somewhat counterintuitive fact that for a realistic array (\emph{i.e.} one with instrumental noise), the errors in the recovered gain parameters may \emph{increase} as one corrects to increasing order in baseline position error $\delta \mathbf{b}_{ij}$.  This is because the higher the order that one corrects to, the greater the number of parameters that one must solve for.  The number of correlation measurements, however, remains the same as before (with $N(N-1)/2$ of them), so the system of equations becomes less constrained.  This makes the calibration algorithm more susceptible to noise, which can negate the benefit that one obtains by correcting baseline position errors to higher order.  Put another way, when we introduce a large number of parameters in an attempt to correct for baseline position errors, we run the risk of over-fitting the instrumental noise instead of averaging it down using redundant baselines.

In running the baseline error simulations, we find the errors in recovered $\varphi$ to be much worse than those in the recovered $\eta$'s.  This is likely due to the fact that a small perturbation in antenna position can result in large changes in phases, which (as explained in Section \ref{logprobs}) can cause problems in a logarithmic implementation.

The reader should also note that because quantities like $\mathbf{h}_0^{\alpha}$ are dependent on the sky signal, we find that the calibration errors induced by baseline position errors cannot be easily captured by simple formulae analogous to Equation \ref{neateqn}.  An extreme example of this was already discussed above in Section \ref{zero}, where we saw that the correction terms for baseline errors approach zero as the sky becomes increasingly dominated by a single bright point source.  The values on the vertical axis of Figure \ref{differentorders} are thus not generically applicable; in general, one must simulate the relevant array arrangement and sky signal to estimate the errors incurred by antenna position errors.

\subsection{First Order: Non-coplanar Arrays}
The methods described above can also be used to calibrate radio interferometers even if there are slight deviations from planarity.  Suppose the antenna elements of our array are all almost (but not quite) located at a height of $z=0$.  The slight deviations from perfect coplanarity of the antenna elements result in slight deviations from coplanarity of the baselines, which we denote $\delta b_z^{ij}$.  With these deviations, Equation \ref{basiceqn} becomes
\begin{equation}
\label{noncoplanarbasiceqn}
c(\mathbf{b}) = \int_{sky} g_i^* g_j \frac{J (\boldsymbol{\theta}) B(\boldsymbol{\theta})}{\sqrt{1-\theta_x^2 - \theta_y^2}} \exp \left( -\frac{i 2 \pi \mathbf{b} \cdot \boldsymbol{\theta}}{\lambda} -\frac{i 2 \pi}{\lambda} \delta b_{z}^{ij} \sqrt{1-\theta_x^2 - \theta_y^2}  \right) d^2 \boldsymbol{\theta}.
\end{equation}
Like before, this equation can be manipulated into a form conducive to calibration if one (or both) of the following conditions are met:
\begin{enumerate}
\item \textbf{Narrow primary beam}.  With a narrow primary beam, one has $\sqrt{1-\theta_x^2 - \theta_y^2} \approx 1- \frac{\theta_x^2}{2} - \frac{\theta_y^2}{2} \dots \approx 1$.  The part of the expression corresponding to the non-coplanar correction thus factors out of the integral, giving
\begin{equation}
c(\mathbf{b}) = \exp\left( -\frac{i 2 \pi}{\lambda} \delta b_{z}^{ij} \right) \int_{sky} g_i^* g_j \frac{J (\boldsymbol{\theta}) B(\boldsymbol{\theta})}{\sqrt{1-\theta_x^2 - \theta_y^2}} \exp \left( -\frac{i 2 \pi \mathbf{b} \cdot \boldsymbol{\theta}}{\lambda}  \right) d^2 \boldsymbol{\theta} =  \exp\left( -\frac{i 2 \pi}{\lambda} \delta b_{z}^{ij} \right) g_i^* g_j \tilde{I} \left( \frac{\mathbf{b}}{\lambda} \right),
 \label{noncoplanarphaseshift}
\end{equation}
where we have adopted the notation defined in Equation \ref{deffourier}.  This illustrates the fact that so long as the primary beam is narrow, deviations from coplanarity simply results in phase shifts in the measured correlations.  Aside from these phase shifts, Equation \ref{noncoplanarphaseshift} looks completely identical to Equation \ref{fourier}, and thus the calibration of a non-coplanar array is no harder than the calibration of a planar array when the primary beam is narrow.  One simply defines adjusted correlations $\tilde{c} \equiv c  \exp\left( -i \frac{2 \pi}{\lambda} \delta b_{z}^{ij} \right) $ and the ensuing analysis proceeds like before, with no additional computational cost.
\item \textbf{Near coplanar array.}  If the array is close to coplanar ($\delta b_z^{ij} \ll 1$), the calibration can still be performed, but this time at a slight increase in computational cost.  With near coplanarity, Equation \ref{noncoplanarbasiceqn} can be Taylor expanded in $\delta b_z$:
\begin{eqnarray}
c(\mathbf{b}) &=& \int_{sky} g_i^* g_j \frac{J (\boldsymbol{\theta}) B(\boldsymbol{\theta})}{\sqrt{1-\theta_x^2 - \theta_y^2}} \exp \left( -\frac{i 2 \pi \mathbf{b} \cdot \boldsymbol{\theta}}{\lambda} \right) \left( 1-\frac{i 2 \pi}{\lambda} \delta b_{z}^{ij} \sqrt{1-\theta_x^2 - \theta_y^2}  \right)d^2 \boldsymbol{\theta} \\
&=&  g_i^* g_j \tilde{I} \left( \frac{\mathbf{b}}{\lambda} \right) - i g_i^* g_j \frac{2 \pi}{\lambda} \delta b_z^{ij} \tilde{K} \left( \frac{\mathbf{b}}{\lambda} \right),
\end{eqnarray}
where $\tilde{I}$ is defined as before in Equations \ref{basiceqn} and \ref{deffourier} and
\begin{equation}
\tilde{K} (\mathbf{k}) \equiv  \int J (\boldsymbol{\theta}) B(\boldsymbol{\theta}) \exp ( -i 2 \pi \mathbf{k} \cdot \boldsymbol{\theta} ) d^2 \boldsymbol{\theta}.
\end{equation}
Perturbing this equation around perfectly redundant baselines using exactly the same methods as those employed between Equations \ref{taylor1} and \ref{linearfirstorder}, one obtains
\begin{equation}
c(\mathbf{b}) = g_i^* g_j \left[ c_0^{\alpha} \left( 1+ \mathbf{h}_0^{\alpha} \cdot \frac{\delta \mathbf{b}_{ij}}{\lambda}\right) -i 2 \pi \frac{\delta b_z^{ij}}{\lambda}d_0^{\alpha}\right],
\end{equation}
where $d_0^{\alpha} \equiv \tilde{K} \left( \frac{\mathbf{b}_0^{\alpha}}{ \lambda} \right)$.  Note that whereas $\tilde{I}$ is expanded to \emph{first} order in $\delta \mathbf{b}$, we are only required to keep the zeroth order term for $\tilde{K}$ since any linear term in $\delta \mathbf{b}$ would be multiplied by $\delta b_z$ and end up a second order term.  Proceeding as before, the analog of Equation \ref{firstexp} takes the form
\begin{equation}
\ln c_{ij} \approx (\eta_i + \eta_j) + i (\varphi_j - \varphi_i )  + \ln c_0^{\alpha} + \mathbf{h}_0^{\alpha} \cdot \frac{\delta \mathbf{b}_{ij}}{\lambda} + i 2\pi q_0^{\alpha} \frac{\delta b_z^{ij}}{\lambda},
\end{equation}
where $q_0^{\alpha} \equiv \ln (d_0^{\alpha} / c_0^{\alpha} )$.  From here, we can once again form a linear system of equations that can be solved to yield the calibration parameters, the only difference being that we must solve for yet one more parameter per cluster of baselines.
\end{enumerate}
\section{Conclusions}
\label{conc}
Redundant calibration schemes calibrate radio interferometers by taking advantage of the fact that --- once calibrated --- redundant sets of baselines should yield identical visibility measurements regardless of what the sky looks like.  In this paper we have performed simulations to examine the error properties of redundant calibration, and have found that the usual logarithmic implementations are statistically biased.  The linearized implementation introduced in Section \ref{linear}, on the other hand, is both unbiased and computationally feasible, being far less computationally intensive than the standard initial step of computing the signal correlations.

Though for many of our simulations we deliberately chose unfavorable SNRs to exaggerate the effects of instrumental noise, the errors are seen to scale inversely with SNR, which means that Equation \ref{neateqn} and Figure \ref{yerrors} predict that redundant baseline calibration can yield high precision calibration parameters in a realistic application.  Moreover, the fact that the linearized method of Section \ref{linear} is unbiased means that even if the SNR \emph{were} unfavorable, one could simply build larger arrays and average over long time periods, in principle suppressing calibration errors to arbitrarily low levels.

We have also shown that both point source calibration and redundant calibration can be considered special cases within a generalized framework of algorithms, all of which find ways to reduce the number of calibration and visibility parameters so that there are few enough of them to be solved for using the $N(N-1)/2$ measurement equations.  This can be done either by making \emph{a priori} assumptions about the calibration sources, or by taking advantage of baseline redundancy.  If one has many more constraints than unknowns, non-exact redundancy and non-coplanarity can also be taken into account, which may be essential as the precision requirements for calibration become more and more stringent.  That redundant baseline calibration appears able to satisfy such requirements while being computationally feasible is encouraging for large radio arrays, suggesting that calibration will not limit the vast scientific potential of $21\,\textrm{cm}$ tomography.

\section*{Acknowledgments}
We wish to thank Judd Bowman, Michiel Brentjens, Joshua Dillon, Jacqueline Hewitt, Miguel Morales, Ue-Li Pen, Kris Sigurdson, Leo Stein, Christopher Williams, and Kristian Zarb Adami for useful discussions.  This work was supported by NSF grants AST-0506556, AST-0907969, AST-0708534, AST-6920223, and PHY-0855425 and NASA NN-6056J406 in addition to fellowships from the David and Lucile Packard Foundation, the Research Corporation and the John and Catherine MacArthur Foundation.

\bibliographystyle{mn2e}
\bibliography{autocalib}

\begin{thebibliography}{}

\bibitem[\protect\citeauthoryear{{Barkana} \& {Loeb}}{{Barkana} \&
  {Loeb}}{2005}]{Barkana1}
{Barkana} R.,  {Loeb} A.,  2005, Ap. J., 626, 1

\bibitem[\protect\citeauthoryear{Bhatnagar, Cornwell, Golap \& Uson}{Bhatnagar
  et~al.}{2008}]{direcdep}
Bhatnagar S.,  Cornwell T.~J.,  Golap K.,    Uson J.~M.,  2008, A\&A, 487, 419

\bibitem[\protect\citeauthoryear{{Bowman}, {Morales} \& {Hewitt}}{{Bowman}
  et~al.}{2007}]{juddjackiemiguel1}
{Bowman} J.~D.,  {Morales} M.~F.,    {Hewitt} J.~N.,  2007, Ap. J., 661, 1

\bibitem[\protect\citeauthoryear{{Carilli} \& {Rawlings}}{{Carilli} \&
  {Rawlings}}{2004}]{SKA}
{Carilli} C.~L.,  {Rawlings} S.,  2004, Science with the Square Kilometer
  Array.
Elsevier, Amsterdam

\bibitem[\protect\citeauthoryear{{Chang}, {Pen}, {Peterson} \&
  {McDonald}}{{Chang} et~al.}{2008}]{ChangDE}
{Chang} T.,  {Pen} U.,  {Peterson} J.~B.,    {McDonald} P.,  2008, Phys. Rev.
  Lett., 100, 091303

\bibitem[\protect\citeauthoryear{{Cornwell} \& {Fomalont}}{{Cornwell} \&
  {Fomalont}}{1999}]{review1}
{Cornwell} T.,  {Fomalont} E.~B.,  1999, in {G.~B.~Taylor, C.~L.~Carilli, \&
  R.~A.~Perley} ed., Synthesis Imaging in Radio Astronomy II Vol.~180 of
  Astronomical Society of the Pacific Conference Series, {Self-Calibration}.
pp 187--+

\bibitem[\protect\citeauthoryear{{Datta}, {Bhatnagar} \& {Carilli}}{{Datta}
  et~al.}{2009}]{Datta}
{Datta} A.,  {Bhatnagar} S.,    {Carilli} C.~L.,  2009, Ap. J., 703, 1851

\bibitem[\protect\citeauthoryear{{de Oliveira-Costa}, {Tegmark}, {Gaensler},
  {Jonas}, {Landecker} \& {Reich}}{{de Oliveira-Costa}
  et~al.}{2008}]{angelicaGSM}
{de Oliveira-Costa} A.,  {Tegmark} M.,  {Gaensler} B.~M.,  {Jonas} J.,
  {Landecker} T.~L.,    {Reich} P.,  2008, MNRAS, 388, 247

\bibitem[\protect\citeauthoryear{{Furlanetto}, {Lidz}, {Loeb}, {McQuinn},
  {Pritchard}, {Alvarez}, {Backer}, {Bowman}, {Burns}, {Carilli}, {Cen},
  {Cooray}, {Gnedin}, {Greenhill}, {Haiman}, {Hewitt} \& et al.}{{Furlanetto}
  et~al.}{2009}]{Whitepaper1}
{Furlanetto} S.~R.,  {Lidz} A.,  {Loeb} A.,  {McQuinn} M.,  {Pritchard} J.~R.,
  {Alvarez} M.~A.,  {Backer} D.~C.,  {Bowman} J.~D.,  {Burns} J.~O.,  {Carilli}
  C.~L.,  {Cen} R.,  {Cooray} A.,  {Gnedin} N.,  {Greenhill} L.~J.,  {Haiman}
  Z.,  {Hewitt} J.~N.,    et al. 2009, in AGB Stars and Related
  Phenomenastro2010: The Astronomy and Astrophysics Decadal Survey Vol.~2010 of
  Astronomy, {Astrophysics from the Highly-Redshifted 21 cm Line}.
pp 83--+

\bibitem[\protect\citeauthoryear{{Furlanetto}, {Lidz}, {Loeb}, {McQuinn},
  {Pritchard}, {Shapiro}, {Alvarez}, {Backer}, {Bowman}, {Burns}, {Carilli},
  {Cen}, {Cooray}, {Gnedin}, {Greenhill}, {Haiman}, {Hewitt} \& et
  al.}{{Furlanetto} et~al.}{2009}]{Whitepaper2}
{Furlanetto} S.~R.,  {Lidz} A.,  {Loeb} A.,  {McQuinn} M.,  {Pritchard} J.~R.,
  {Shapiro} P.~R.,  {Alvarez} M.~A.,  {Backer} D.~C.,  {Bowman} J.~D.,  {Burns}
  J.~O.,  {Carilli} C.~L.,  {Cen} R.,  {Cooray} A.,  {Gnedin} N.,  {Greenhill}
  L.~J.,  {Haiman} Z.,  {Hewitt} J.~N.,    et al. 2009, in AGB Stars and
  Related Phenomenastro2010: The Astronomy and Astrophysics Decadal Survey
  Vol.~2010 of Astronomy, {Cosmology from the Highly-Redshifted 21 cm Line}.
pp 82--+

\bibitem[\protect\citeauthoryear{{Furlanetto}, {Sokasian} \&
  {Hernquist}}{{Furlanetto} et~al.}{2004}]{furlanetto1}
{Furlanetto} S.~R.,  {Sokasian} A.,    {Hernquist} L.,  2004, MNRAS, 347, 187

\bibitem[\protect\citeauthoryear{{Furlanetto}, {Zaldarriaga} \&
  {Hernquist}}{{Furlanetto} et~al.}{2004}]{furlanetto2}
{Furlanetto} S.~R.,  {Zaldarriaga} M.,    {Hernquist} L.,  2004, Ap. J., 613,
  16

\bibitem[\protect\citeauthoryear{{Garrett}}{{Garrett}}{2009}]{LOFAR}
{Garrett} M.~A.,  2009, ArXiv e-prints: 0909.3147 (astro-ph)

\bibitem[\protect\citeauthoryear{{Hinshaw}, {Barnes}, {Bennett}, {Greason},
  {Halpern}, {Hill}, {Jarosik}, {Kogut}, {Limon}, {Meyer}, {Odegard}, {Page},
  {Spergel}, {Tucker}, {Weiland}, {Wollack} \& {Wright}}{{Hinshaw}
  et~al.}{2003}]{WMAP}
{Hinshaw} G.,  {Barnes} C.,  {Bennett} C.~L.,  {Greason} M.~R.,  {Halpern} M.,
  {Hill} R.~S.,  {Jarosik} N.,  {Kogut} A.,  {Limon} M.,  {Meyer} S.~S.,
  {Odegard} N.,  {Page} L.,  {Spergel} D.~N.,  {Tucker} G.~S.,  {Weiland}
  J.~L.,  {Wollack} E.,    {Wright} E.~L.,  2003, ApJS, 148, 63

\bibitem[\protect\citeauthoryear{{Iliev}, {Shapiro}, {Ferrara} \&
  {Martel}}{{Iliev} et~al.}{2002}]{Iliev}
{Iliev} I.~T.,  {Shapiro} P.~R.,  {Ferrara} A.,    {Martel} H.,  2002, ApJL,
  572, L123

\bibitem[\protect\citeauthoryear{{Ishiguro}}{{Ishiguro}}{1974}]{oldredund2}
{Ishiguro} M.,  1974, Astronomy and Astrophysics Supplement, 15, 431

\bibitem[\protect\citeauthoryear{{Lidz}, {Zahn}, {McQuinn}, {Zaldarriaga} \&
  {Hernquist}}{{Lidz} et~al.}{2008}]{adam}
{Lidz} A.,  {Zahn} O.,  {McQuinn} M.,  {Zaldarriaga} M.,    {Hernquist} L.,
  2008, ApJ, 680, 962

\bibitem[\protect\citeauthoryear{{Liu}, {Tegmark} \& {Zaldarriaga}}{{Liu}
  et~al.}{2009}]{us}
{Liu} A.,  {Tegmark} M.,    {Zaldarriaga} M.,  2009, MNRAS, 394, 1575

\bibitem[\protect\citeauthoryear{{Loeb} \& {Zaldarriaga}}{{Loeb} \&
  {Zaldarriaga}}{2004}]{Loeb1}
{Loeb} A.,  {Zaldarriaga} M.,  2004, Physical Review Letters, 92, 211301

\bibitem[\protect\citeauthoryear{{Lonsdale}, {Cappallo}, {Morales}, {Briggs},
  {Benkevitch}, {Bowman}, {Bunton}, {Burns}, {Corey}, {Desouza}, {Doeleman},
  {Derome}, {Deshpande}, {Gopala}, {Greenhill}, {Herne}, {Hewitt} \& et
  al.}{{Lonsdale} et~al.}{2009}]{MWAdesign}
{Lonsdale} C.~J.,  {Cappallo} R.~J.,  {Morales} M.~F.,  {Briggs} F.~H.,
  {Benkevitch} L.,  {Bowman} J.~D.,  {Bunton} J.~D.,  {Burns} S.,  {Corey}
  B.~E.,  {Desouza} L.,  {Doeleman} S.~S.,  {Derome} M.,  {Deshpande} A.,
  {Gopala} M.~R.,  {Greenhill} L.~J.,  {Herne} D.~E.,  {Hewitt} J.~N.,    et
  al. 2009, IEEE Proceedings, 97, 1497

\bibitem[\protect\citeauthoryear{{Madau}, {Meiksin} \& {Rees}}{{Madau}
  et~al.}{1997}]{Rees}
{Madau} P.,  {Meiksin} A.,    {Rees} M.~J.,  1997, Ap. J., 475, 429

\bibitem[\protect\citeauthoryear{{Mao}, {Tegmark}, {McQuinn}, {Zaldarriaga} \&
  {Zahn}}{{Mao} et~al.}{2008}]{Yi}
{Mao} Y.,  {Tegmark} M.,  {McQuinn} M.,  {Zaldarriaga} M.,    {Zahn} O.,  2008,
  PRD, 78, 023529

\bibitem[\protect\citeauthoryear{{McQuinn}, {Zahn}, {Zaldarriaga}, {Hernquist}
  \& {Furlanetto}}{{McQuinn} et~al.}{2006}]{Matt3}
{McQuinn} M.,  {Zahn} O.,  {Zaldarriaga} M.,  {Hernquist} L.,    {Furlanetto}
  S.~R.,  2006, Ap. J., 653, 815

\bibitem[\protect\citeauthoryear{{Morales}}{{Morales}}{2005}]{anothermiguel}
{Morales} M.~F.,  2005, ApJ, 619, 678

\bibitem[\protect\citeauthoryear{{Morales} \& {Matejek}}{{Morales} \&
  {Matejek}}{2009}]{softwareholography}
{Morales} M.~F.,  {Matejek} M.,  2009, MNRAS, 400, 1814

\bibitem[\protect\citeauthoryear{{Morales} \& {Wyithe}}{{Morales} \&
  {Wyithe}}{2009}]{miguelreview}
{Morales} M.~F.,  {Wyithe} J.~S.~B.,  2009, ArXiv e-prints: 0910.3010

\bibitem[\protect\citeauthoryear{{Noordam} \& {de Bruyn}}{{Noordam} \& {de
  Bruyn}}{1982}]{oldredund1}
{Noordam} J.~E.,  {de Bruyn} A.~G.,  1982, Nature, 299, 597

\bibitem[\protect\citeauthoryear{Parsons, Backer, Foster, Wright, Bradley,
  Gugliucci, Parashare, Benoit, Aguirre, Jacobs, Carilli, Herne, Lynch, Manley
  \& Werthimer}{Parsons et~al.}{2010}]{PAPER}
Parsons A.~R.,  Backer D.~C.,  Foster G.~S.,  Wright M. C.~H.,  Bradley R.~F.,
  Gugliucci N.~E.,  Parashare C.~R.,  Benoit E.~E.,  Aguirre J.~E.,  Jacobs
  D.~C.,  Carilli C.~L.,  Herne D.,  Lynch M.~J.,  Manley J.~R.,    Werthimer
  D.~J.,  2010, The Astronomical Journal, 139, 1468

\bibitem[\protect\citeauthoryear{{Pearson} \& {Readhead}}{{Pearson} \&
  {Readhead}}{1984}]{oldredund6}
{Pearson} T.~J.,  {Readhead} A.~C.~S.,  1984, Annu. Rev. of Astronomy and
  Astrophysics, 22, 97

\bibitem[\protect\citeauthoryear{{Peterson}, {Bandura} \& {Pen}}{{Peterson}
  et~al.}{2006}]{CRT}
{Peterson} J.~B.,  {Bandura} K.,    {Pen} U.~L.,  2006, ArXiv Astrophysics
  e-prints: astro-ph/0606104

\bibitem[\protect\citeauthoryear{{Ramesh}, {Subramanian} \& {Sastry}}{{Ramesh}
  et~al.}{1999}]{oldredund3}
{Ramesh} R.,  {Subramanian} K.~R.,    {Sastry} C.~V.,  1999, Astronomy and
  Astrophysics Supplement, 139, 179

\bibitem[\protect\citeauthoryear{{Rau}, {Bhatnagar}, {Voronkov} \&
  {Cornwell}}{{Rau} et~al.}{2009}]{review2}
{Rau} U.,  {Bhatnagar} S.,  {Voronkov} M.~A.,    {Cornwell} T.~J.,  2009, IEEE
  Proceedings, 97, 1472

\bibitem[\protect\citeauthoryear{{Santos} \& {Cooray}}{{Santos} \&
  {Cooray}}{2006}]{Santos2}
{Santos} M.~G.,  {Cooray} A.,  2006, PRD, 74, 083517

\bibitem[\protect\citeauthoryear{{Tegmark}}{{Tegmark}}{1997}]{MaxMaps}
{Tegmark} M.,  1997, Ap. J. Lett., 480, L87+

\bibitem[\protect\citeauthoryear{{Tegmark} \& {Zaldarriaga}}{{Tegmark} \&
  {Zaldarriaga}}{2009a}]{FFTT1}
{Tegmark} M.,  {Zaldarriaga} M.,  2009a, PRD, 79, 083530

\bibitem[\protect\citeauthoryear{{Tegmark} \& {Zaldarriaga}}{{Tegmark} \&
  {Zaldarriaga}}{2009b}]{FFTT2}
{Tegmark} M.,  {Zaldarriaga} M.,  2009b, ArXiv e-prints: 0909.0001

\bibitem[\protect\citeauthoryear{{Tozzi}, {Madau}, {Meiksin} \& {Rees}}{{Tozzi}
  et~al.}{2000a}]{Tozzi2}
{Tozzi} P.,  {Madau} P.,  {Meiksin} A.,    {Rees} M.~J.,  2000a, Ap. J., 528,
  597

\bibitem[\protect\citeauthoryear{{Tozzi}, {Madau}, {Meiksin} \& {Rees}}{{Tozzi}
  et~al.}{2000b}]{Tozzi}
{Tozzi} P.,  {Madau} P.,  {Meiksin} A.,    {Rees} M.~J.,  2000b, Nuclear
  Physics B Proceedings Supplements, 80, C509+

\bibitem[\protect\citeauthoryear{{Wieringa}}{{Wieringa}}{1992}]{oldredund4}
{Wieringa} M.~H.,  1992, Experimental Astronomy, 2, 203

\bibitem[\protect\citeauthoryear{{Wyithe}, {Loeb} \& {Geil}}{{Wyithe}
  et~al.}{2008}]{wyithe2008}
{Wyithe} J.~S.~B.,  {Loeb} A.,    {Geil} P.~M.,  2008, MNRAS, 383, 1195

\bibitem[\protect\citeauthoryear{{Yang}}{{Yang}}{1988}]{oldredund5}
{Yang} Y.,  1988, Astronomy and Astrophysics, 189, 361

\end{thebibliography}

\end{document}